\documentclass[twocolumn,aps,pra,showpacs,superscriptaddress,floatfix]{revtex4}
\usepackage{graphicx}
\usepackage{times}
\usepackage{nicefrac}
\usepackage{amsmath}
\usepackage{amsfonts}
\usepackage{amssymb}
\usepackage{amsthm}
\usepackage{epsf}
\usepackage{bm}
\usepackage{bbm}

\usepackage{dcolumn}
\newcolumntype{.}{D{x}{}{-1}}

\newcommand{\bsigma}{\vec{\sigma}}

\newcommand{\bnabla}{\vec{\nabla}}
\newcommand{\bpi}{\vec{\pi}}
\newcommand{\bfr}{\vec{r}}
\newcommand{\bfk}{\vec{k}}
\newcommand{\bfp}{\vec{p}}
\newcommand{\bfq}{\vec{q}}

\newcommand{\bfE}{\vec{E}}
\newcommand{\bfA}{\vec{A}}
\newcommand{\bfB}{\vec{B}}

\newcommand{\lbr}{\left<}
\newcommand{\rbr}{\right>}

\newcommand{\cross}[1]{#1\!\!\!/}

\begin{document}

\title{Reexamination of helium fine structure}

\author{Krzysztof Pachucki}
\affiliation{Institute of Theoretical Physics, University of Warsaw,
Ho\.{z}a 69, 00--681 Warsaw, Poland}

\author{Vladimir A. Yerokhin}
\affiliation{Center for Advanced Studies, St.~Petersburg State
Polytechnical University, Polytekhnicheskaya 29, 
St.~Petersburg 195251, Russia}

\begin{abstract}
In order to explain discrepancies between theoretical predictions
and experimental data for the helium fine structure,
we check and recalculate all theoretical contributions up to
orders $m\,\alpha^7$ and $m^2/M\,\alpha^6$.
The previous result for the $m\,\alpha^7$ correction is improved
by a much more accurate calculation of relativistic 
corrections to the Bethe logarithm. 
The theoretical values of the $2^3P_0-2^3P_1$ and $2^3P_1-2^3P_2$
fine structure intervals in helium are, correspondingly,
$\nu_{01} = 29~616~946.2(1.6)$~kHz and $\nu_{12} = 2~291~177.3(1.6)$~kHz, with
the uncertanties being due to higher-order effects. 
For the small interval $\nu_{12}$, the theoretical value agrees with the experimental
data, whereas for the large interval $\nu_{01}$, a discrepancy of about 3 standard
deviations is present.
\end{abstract}


\maketitle

\section{Introduction}

The fine structure splitting of the $2^3P$ level in helium has long been
an attractive subject of theoretical and experimental studies. One of the
reasons for this interest is that the fine structure, being an intrinsically
relativistic effect, is proportional to $\alpha^2$Ry and thus provides an
opportunity for the determination of the fine structure constant $\alpha$ from
a comparison of theoretical predictions with experimental data. A series of 
measurements of the helium fine structure has been
performed during the last decade
\cite{minardi:99,castillega:00,storry:00,george:01,zelevinsky:05,giusfredi:05,borbely:09}, with the
current accuracy being on the level of 25 ppb. For theory, to reach an
adequate level of precision in a description of a three-body system 
is a challenging problem. 

Despite considerable calculational efforts of last years
to provide an accurate theoretical
determination of the fine structure of helium, the current 
status of theory can hardly be considered as satisfactory. Recent
calculations \cite{drake:02:cjp,pachucki:06:prl:he}
demonstrated a significant discrepancy with the
experimental data, the difference for the large (small) fine structure interval being about 
10 (6) times larger than the total nonlogarithmic contribution to order $m\,\alpha^7$. 
It seems unlikely that such
difference can be explained only by higher-order effects. 

The theory of the helium fine structure up to order $m\,\alpha^6$ has been
confirmed by at least two independent calculations and thus can be considered
as established. The logarithmic part of the $m\,\alpha^7$ contribution has
also been calculated independently. The only corrections that are not yet
checked by different evaluations are the recoil contribution to order $m^2/M\,\alpha^6$ and the
nonlogarithmic correction to order $m\,\alpha^7$. These corrections will be the
main subject of the present investigation.

The theoretical description of the fine structure to order
$m\,\alpha^7$ is a difficult task. Within 
the logarithmic accuracy, this was first done
by Zhang {\em et al.} \cite{zhang:96:prl} and later confirmed by one of the authors (K.P.) 
\cite{pachucki:99:jpb}. An
important part of the nonlogarithmic correction to order $m\,\alpha^7$ was
calculated by K.P. and Sapirstein \cite{pachucki:00:jpb}. The problem of derivation of the
complete set of effective spin-dependent operators to order $m\,\alpha^7$ was
addressed by Zhang in a series of works \cite{zhang:96:a,zhang:96:b,zhang:97}
within the equal-time variant of the Bethe-Salpeter formalism. The
derivation based on the dimensionally regularized NRQED was reported recently by 
K.P.~\cite{pachucki:06:prl:he}, who noted several mistakes and inconsistencies
in the previous derivation by Zhang.

In the present investigation, we give a detailed account of the derivation of
the total contribution to order $m\,\alpha^7$, first reported in
Ref.~\cite{pachucki:06:prl:he}, and present a recalculation of all corrections
up to orders $m^2/M\,\alpha^6$ and $m\,\alpha^7$. Particularly, we perform
an evaluation of the relativistic correction to the Bethe logarithm, which
improves significantly upon the first calculation in Ref. \cite{pachucki:00:jpb}. 

The paper is organized as follows. In Sec.~\ref{sec1} and \ref{sec2} we give
a short summary of the general formulas for the helium fine structure up to
order $m\,\alpha^6$. The derivation of the $m\,\alpha^7$ correction 
is presented in Sec.~\ref{sec3}.
The next section describes the numerical approach and reports the numerical
results. Sec.~\ref{sec:summary} contains the summary of all contributions to 
the helium fine structure and
the discussion of the present status of theory and experiment. 

The relativistic units are used in this paper, $\hbar = c =
\epsilon_0 = 1$ and $e^2 = 4\pi\alpha$.

\section{Leading-order fine structure}
\label{sec1}

The dominant contribution to the helium fine structure is induced by the
spin dependent part of the Breit-Pauli Hamiltonian, 
which is, for an infinitely heavy nucleus,
\begin{eqnarray} \label{fs}
H_{\rm fs} & = &\frac{\alpha}{4\,m^2}\left(
\frac{\bsigma_1\cdot\bsigma_2}{r^3}
-3\,\frac{\bsigma_1\cdot\bfr\,
\bsigma_2\cdot\bfr}{r^5}\right)(1+a_e)^2\, \nonumber \\
& + & {Z\alpha  \over 4 m^2} 
\left[
\frac{1}{r_1^3}\,\bfr_1\times\bfp_1\cdot\bsigma_1+
\frac{1}{r_2^3}\,\bfr_2\times\bfp_2\cdot\bsigma_2
\right](1+2a_e)  
\nonumber \\
& + & 
\frac{\alpha}{4\,m^2
\,r^3}\biggl[
\bigl[(1+2\,a_e)\,\bsigma_2+2\,(1+a_e)\,\bsigma_1\bigr]\cdot\bfr\times\bfp_2
\nonumber \\ &&
-\bigl[(1+2\,a_e)\,\bsigma_1+2\,(1+a_e)\,\bsigma_2\bigr]\cdot\bfr
\times\bfp_1\biggr]\,,
\label{HSD}
\end{eqnarray}
where $\bfr = \bfr_1-\bfr_2$ and we have
included the effects of the electron anomalous magnetic moment
(amm) $a_e$,
\begin{eqnarray}
a_e &=& \frac{\alpha}{2\pi} -0.328\,478\,965\,\left(\frac{\alpha}{\pi}\right)^2
\nonumber \\ &&
  +1.181\,241\,456\,\left(\frac{\alpha}{\pi}\right)^3
  -1.7283\,(35)\,\left(\frac{\alpha}{\pi}\right)^4 +\ldots\,.
\end{eqnarray} 
Expanding the amm prefactors in Eq.~(\ref{fs}), $H_{\rm fs}$ can be written as a
sum of operators contributing to different orders in $\alpha$,
\begin{eqnarray} \label{fs2}
H_{\rm fs} = H_{\rm fs}^{(4)}+ H_{\rm fs}^{(5)}+ H_{\rm fs,amm}^{(6)}
+ H_{\rm fs,amm}^{(7)}+ \ldots\,.
\end{eqnarray} 
Here $H_{\rm fs}^{(4)}$ and $H_{\rm fs}^{(5)}$ yield the complete 
fine-structure contributions 
of order $m\,\alpha^4$ and $m\,\alpha^5$, respectively,
whereas $H_{\rm fs,amm}^{(6)}$ and $H_{\rm fs,amm}^{(7)}$ are the amm parts of the
corresponding higher-order operators.
 
The leading effect of the finite nuclear mass is
conveniently divided into three parts, termed as the mass scaling, the
mass polarization, and the recoil operators. 
The effect of the mass scaling is accounted
for by including the prefactor $(m_r/m)^3$ into the operator $H_{\rm fs}$,
where $m_r$ is the reduced mass for the electron-nucleus system. The effect of
the mass polarization can be accounted for to all orders by evaluating
expectation values of all operators on the 
eigenfunctions of the Shr\"odinger Hamiltonian with the mass-polarization
operator $(m_r/M)\, \bfp_1\cdot\bfp_2$ included. The third effect is
induced by the recoil addition to the Breit-Pauli Hamiltonian, 
\begin{eqnarray} \label{fsrec}
H_{\rm fs,rec} &=& \frac{Z\alpha}{2mM}\, \left[
\frac{\bfr_1}{r_1^3}\times(\bfp_1+\bfp_2)\cdot\bsigma_1
\right.
\nonumber \\ &&
\left.
+ \frac{\bfr_2}{r_2^3}\times(\bfp_1+\bfp_2)\cdot\bsigma_2 
\right]
(1+a_e)\,.
\end{eqnarray}

\section{$\bm{m}\, \bm{\alpha}^{\bf 6}$ contribution}
\label{sec2}

The $m\,\alpha^6$ contribution to the helium fine structure is a sum of the
second-order perturbation corrections induced by the Breit-Pauli Hamiltonian
and the expectation value of 
the effective fine-structure Hamiltonian to this order, $H^{(6)}_{\rm fs}$, 
\begin{eqnarray}  \label{fs6}
E^{(6)} &=& \lbr H_{\rm fs} \frac1{(E_0-H_0)'} H_{\rm fs} \rbr
\nonumber \\ & & 
   + 2\,\lbr H^{(4)}_{\rm nfs} \frac1{(E_0-H_0)'} H_{\rm fs} \rbr
    + \lbr H^{(6)}_{\rm fs} \rbr\,.
\end{eqnarray} 
Here, $1/(E_0-H_0)'$ is the reduced Green function and
$H^{(4)}_{\rm nfs}$ is the spin-independent part of the Breit-Pauli
Hamiltonian, 
\begin{eqnarray} \label{nfs}
H^{(4)}_{\rm nfs} & = & -\frac{1}{8\,m^3}\,(p_1^4+p_2^4)+
\frac{Z\,\alpha\,\pi}{2\,m^2}\,\bigl[\delta^3(r_1)+\delta^3(r_2)\bigr]
\nonumber \\ & & 
-\frac{\alpha}{2\,m^2}\,p_1^i\,
\biggl(\frac{\delta^{ij}}{r}+\frac{r^i\,r^j}{r^3}\biggr)\,p_2^j
\,,
\end{eqnarray}
where we omitted a term with $\delta^3(r)$ since it vanishes for
the triplet states.
It is noteworthy that in Eq.~(\ref{fs6}) we include the operator $H_{\rm fs}$
[and not just $H^{(4)}_{\rm fs}$], thus accounting for the amm correction to
the Breit-Pauli Hamiltonian. While this correction is of order $m\,\alpha^7$, it
is convenient to calculate it together with the $m\,\alpha^6$ contribution
because only simple changes in the prefactors are required. 

$H^{(6)}_{\rm fs}$ consists of 15 operators first derived by Douglas and Kroll (DK)
\cite{douglas:74} in the framework of the Salpeter equation. These operators
were later re-derived in a more simple way using the effective field theory
in Refs.~\cite{zhang:96:a,pachucki:99:jpb}. The result is
\begin{eqnarray} 
   H^{(6)}_{\rm fs}  = \sum_{i=1}^{15} B_i\,,
\end{eqnarray} 
where
\begin{eqnarray}  \label{DK0}
B_1 &=& \frac{3\,Z}{8}\,\nabla_1^2\,\frac{1}{r_1^3}\,\bsigma_1\cdot(\bfr_1\times\bfp_1) \,,\\
B_2 &=& -\frac{Z}{r^3\,r_1^3}\,\bsigma_1\cdot(\bfr_1\times\bfr)\,(\bfr\cdot\bfp_2)           \,,\\
B_3 &=& \frac{Z}{2}\,\,\frac{1}{r^3\,r_1^3}\,(\bsigma_1\cdot\bfr)\,(\bsigma_2\cdot\bfr_1) \,,\\
B_4 &=& \frac{1}{2}\,\frac{1}{r^4}\,\bsigma_1\cdot(\bfr\times\bfp_2) \,,\\
B_5 &=& -\frac{1}{2}\,\frac{1}{r^6}\,(\bsigma_1\cdot\bfr)\,(\bsigma_2\cdot\bfr) \,,\\
B_6 &=& -\frac{5}{8}\,\nabla_1^2\,\frac{1}{r^3}\,\bsigma_1\cdot(\bfr\times\bfp_1) \,,\\
B_7 &=& \frac{3}{4}\,\nabla_1^2\,\frac{1}{r^3}\,\bsigma_1\cdot(\bfr\times\bfp_2) \,,\\
B_8 &=&
\frac{i}{4}\,\nabla_1^2\,\frac{1}{r}\,\bsigma_1\cdot(\bfp_1\times\bfp_2) \,, \label{B8}\\ 
B_9 &=& \frac{3\,i}{4}\,\nabla_1^2\,\frac{1}{r^3}\,(\bfr\cdot\bfp_2)\,\bsigma_1\cdot(\bfr\times\bfp_1) \,,\\
B_{10} &=& \frac{3\,i}{8}\,\frac{1}{r^5}\,\bsigma_1\cdot(\bfr\times(\bfr\cdot\bfp_2)\,\bfp_1) \,,\\
B_{11} &=& -\frac{3}{16}\,\frac{1}{r^5}\,\bsigma_2\cdot(\bfr\times(\bsigma_1\cdot(\bfr\times\bfp_1))\,\bfp_2) \,,\\
B_{12} &=& -\frac{1}{16}\,\frac{1}{r^3}\,(\bsigma_1\cdot\bfp_2)\,(\bsigma_2\cdot\bfp_1) \,,\\
B_{13} &=& -\frac{3}{2}\,\nabla_1^2\,\frac{1}{r^5}\,(\bsigma_1\cdot\bfr)\,(\bsigma_2\cdot\bfr) \,,\\
B_{14} &=& \frac{i}{4}\,\nabla_1^2\,\frac{1}{r^3}\,\,(\bsigma_1\cdot\bfr)\,(\bsigma_2\cdot\bfp_1) \,,\\
B_{15} &=& -\frac{i}{8}\,\nabla_1^2\,\frac{1}{r^3}\,\,(\bsigma_1\cdot\bfr)\,(\bsigma_2\cdot\bfp_2) \,.
\label{DK1}
\end{eqnarray} 

The finite nuclear mass correction correction to the $m\,\alpha^6$
contribution can be divided into the mass scaling, the mass polarization,
and the operator parts. The mass scaling prefactor is $(m_r/M)^4$ for the 
$B_2$, $B_3$, $B_4$, and $B_5$, $(m_r/M)^5$ for the other $B_i$ operators, 
$(m_r/M)^6$ for the second-order corrections involving the first term in
Eq.~(\ref{nfs}), and $(m_r/M)^5$ for all other second-order corrections. The
mass polarization effect is most easily accounted for by including the mass
polarization operator into the zeroth-order Hamiltonian. The operator part 
comes from recoil corrections to $H^{(4)}_{\rm fs}$, $H^{(4)}_{\rm nfs}$,
and $H^{(6)}_{\rm fs}$. The recoil part of $H^{(4)}_{\rm fs}$ is given by
Eq.~(\ref{fsrec}). The spin-independent recoil part of the  Breit-Pauli
Hamiltonian is 
\begin{align}
  H_{\rm nfs,rec}^{(4)} &\ = 
    -\frac{Z}{2}\,\frac{m}{M} \sum_{a=1,2}
    p_a^i\,\left(\frac{\delta^{ij}}{r_a}+\frac{r^i_ar^j_a}{r_a^3}\right)
  (p_1^j+p_2^j)\,.
\end{align}
Recoil corrections to the DK operators were studied by Zhang \cite{zhang:97}
and by K.P. and Sapirstein \cite{pachucki:03:jpb}. The result is given by the
effective Hamiltonian  
\begin{eqnarray} 
  H^{(6)}_{\rm fs,rec}  = \frac{m}{M} \sum_{i=1}^{8} V_i\,,
\end{eqnarray} 
where
\begin{eqnarray} 
V_1 &=& { i Z \over 4 } p_1^2 {1 \over r_1} \bsigma_1 \cdot (\bfp_1 \times \bfp_2) 
  \,,\\
V_2 &=& -\frac{i Z}{4}\,p_1^2\,\frac{\bfr_1}{r_1^3}\,
(\bsigma_1 \cdot \bfr_1\times \bfp_1)\cdot (\bfp_1+\bfp_2) 
 \,,\\
V_3 &=& -\frac{3 Z}{4}\,p_1^2\,\bsigma_1 \cdot \frac{\bfr_1}{r_1^3}\times (\bfp_1+\bfp_2) 
 \,,\\
V_4 &=&  Z \,\bsigma_1 \cdot \frac{\bfr}{r_1\,r^3}\times (\bfp_1+\bfp_2) 
 \,,\\
V_5 &=& Z\, \bsigma_1 \cdot {\bfr \over r^3} \times
{\bfr_{1} \over r_{1}^3}\,(\bfr_1 \cdot(\bfp_1 + \bfp_2))
 \,,\\
V_6 &=&  Z^2\,\bsigma_1 \cdot {\bfr_1 \over r_1^3} \times {\bfr_2 \over r_2^3}\,
(\bfr_1 \cdot \bfp_1) 
 \,,\\
V_7 &=&  -\frac{Z^2}{2} \,\bsigma_1 \cdot \frac{\bfr_{1}}{r_{1}^4}\times (\bfp_1+\bfp_2) 
 \,,\\
V_8 &=& -{Z^2 \over 4} \bsigma_1 \cdot {\bfr_2 \over r_2^3} \,
              \bsigma_2 \cdot {\bfr_1 \over r_1^3}\,.
\end{eqnarray}


\section{Derivation of the $\bm{m}\,\bm{\alpha}^{\bf 7}$ contribution}
\label{sec3}

In this section we present a detailed derivation of the $m\,\alpha^7$
contribution to the helium fine structure. The corresponding results
have already been presented in Ref. \cite{pachucki:06:prl:he}.
The derivation is based on the dimensionally regularized NRQED \cite{yelkhovsky:01:pra}.
The general idea is that, in the situation when all relevant electron momenta
are much smaller than the electron mass, an approximate QED Lagrangian can
be used, obtained from the original full-QED Lagrangian 
by the Foldy-Wouthuysen (FW) transformation, as described in Appendix~\ref{appendix}.
The standard FW transformation is generalized to the extended number of the
space dimensions and also to account for the magnetic moment anomaly of the 
electron $a_e$.
The regularization parameter $\epsilon$, related to the space dimension $d=3-2\,\epsilon$,
plays the role of both an infrared and ultraviolet regulator and cancels out
in the end of calculations.

The fine structure contribution to order $m\,\alpha^7$ ($\alpha^5\,$Ry) 
can be written as \cite{pachucki:00:jpb}
\begin{eqnarray} \label{ma7}
\hspace*{-1ex}E^{(7)}&=& \langle H^{(7)}_{\rm fs}\rangle+
2\left\langle H^{(4)}\frac{1}{(E_0-H_0)'}H^{(5)}\right\rangle+E_{L},
\label{139}
\end{eqnarray} 
where $H^{(i)}$ denotes the effective Hamiltonian to order $m\,\alpha^i$, 
$1/(E_0-H_0)'$ is the reduced Coulomb Green function, and $E_L$ is
the low energy contribution to be interpreted as 
the relativistic correction to the Bethe logarithm. 

The second term in Eq.~(\ref{ma7}) that involves the second-order matrix element
is the simplest. $H^{(4)}$ is the Breit-Pauli
Hamiltonian and is the sum of the spin-dependent and the spin-independent parts,
$H^{(4)} = H^{(4)}_{\rm fs}+ H^{(4)}_{\rm nfs}$.  
The effective Hamiltonian to order $m\, \alpha^5$ is given by the
sum  $H^{(5)} = H^{(5)}_{\rm fs} + H^{(5)}_{\rm nfs}$, where the first part is the
leading-order amm correction to the Breit-Pauli Hamiltonian
defined by Eq.~(\ref{fs2}) and 
\begin{eqnarray} \label{H5nfs}
        H^{(5)}_{\rm nfs}  &=& 
- \frac{7\,\alpha^2}{6\,\pi\,m^2}\,\frac{1}{r^3}
\nonumber \\ && \!\!\!\!
+ \frac{4}{3}\,\frac{Z\alpha^2}{m^2} \,\left[\frac{19}{30} + \ln (Z\alpha)^{-2} \right]
\left[\delta^3(r_1)+\delta^3(r_2)\right] 
\,.\nonumber \\
\end{eqnarray}
The amm part of $H^{(5)}$ has already been included into the $m\,\alpha^6$ correction
described in the previous section. The remaining part of the second-order
perturbation correction will be denoted as $E_S$
and is given by
\begin{eqnarray} \label{ES}
E_S = 2\,\left\langle H^{(4)}_{\rm fs}\,\frac{1}{(E_0-H_0)'}\,H^{(5)}_{\rm nfs}\right\rangle\,.
\end{eqnarray}

The effective operator $H^{(7)}_{\rm fs}$ consists of two parts: (i) the 
exchange terms, in which photons are exchanged between the two electrons, and
(ii) the radiative corrections, in which one or several photons are emitted
and absorbed by the same electron. They are calculated separately 
using different computational methods in the following subsections.

\subsection{Photon exchange part} \label{sec:exchange}

An important feature that leads to a considerable simplification of the
calculation of the photon exchange part
is the fact that the order being calculated is nonanalytic in $\alpha^2$.
For example, $H^{(5)}_{\rm nfs}$ consists of the two terms only, which
can be derived from the two-photon exchange scattering amplitude.
A similar statement holds for the photon exchange part of $H^{(7)}_{\rm fs}$:
if $H^{(7)}_{\rm fs}$ is an effective Hamiltonian, it has to give the same
scattering amplitude as in full QED. Due to the simple 
structure of $H^{(7)}_{\rm fs}$, the effective interaction can be unambiguosly
extracted from this amplitude. The scattering amplitude is usually
much simpler to calculate, than corrections within
effective field approach, such as that used for the derivation of $H^{(6)}$.
An important point here is  that only the two-photon exchange diagrams 
contribute to $H^{(7)}_{\rm fs}$, the absence of the three-body effects 
being a result of an internal cancellation. 

\begin{figure}[htb]%
\begin{center}\includegraphics[width=\columnwidth]{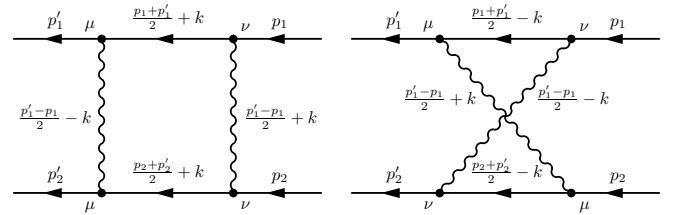}\end{center}%
\caption{\label{fig1}
The two-photon exchange scattering amplitude}%
\end{figure}

So, we obtain the exchange contribution from the spin dependent 
part of the two-photon scattering amplitude, which is
\begin{eqnarray} \label{aa1}
\delta_1 H &=& \frac{i\,e^4}{(2\,\pi)^D}\,\int d^D k\,
\frac{1}{(k+q/2)^2}\,\frac{1}{(k-q/2)^2}\nonumber \nonumber \\ &&
\biggl[\bar u(p'_1)\,\gamma^\mu\,\frac{1}
{\cross{k}+(\cross{p}_1 + \cross{p}'_1)/2-1}\,\gamma^\nu\,u(p_1) \nonumber \\ &&
+\bar u(p'_1)\,\gamma^\nu\,\frac{1}
{-\cross{k}+(\cross{p}_1 + \cross{p}'_1)/2-1}\,\gamma^\mu\,u(p_1)\biggr]
\nonumber \\ &&
\times \bar u(p'_2)\,\gamma^\nu\,\frac{1}
{\cross{k}+(\cross{p}_2 + \cross{p}'_2)/2-1}\,\gamma^\mu\,u(p_2)\,,\label{36}
\end{eqnarray}
where $q=p'_1-p_1$. There are three scales of the $k$ integral
that are responsible for the $m\,\alpha^7$ corrections:
$m$, $m\,\alpha$, and $m\,\alpha^2$.
Only the first two scales are accounted for in Eq.~(\ref{aa1}), whereas
the third one, $k\sim m\,\alpha^2$, corresponds to a low-energy contribution and 
requires a separate treatment. 
Because of the dimensional regularization, the contribution of 
each energy scale can be obtained separately.

In our calculation, only the spin-dependent part of the 
scattering amplitude $\delta_1 H$ is needed. 
It is, however, not obvious what the spin dependent part is.
In order to be consistent with 
the rest of the calculation, we employ the free  
FW transformation $S$,
\begin{eqnarray}
\psi' &=& e^{i\,S}\,,\\
e^{-i\,S} &=& \frac{\cross{p}+1}{\sqrt{2\,E_p\,(E_p+1)}}\,,
\end{eqnarray}
which for small momentum takes a simple form
\begin{equation}
e^{-i\,S} \approx \frac{\cross{p}+1}{2}\,.
\end{equation}
This leads us to the following projection operators 
\begin{align}
\bar u(p')\,Q\,u(p) &\ = {\rm Tr}\,Q\,u(p)\otimes\bar u(p') 
 \nonumber \\ &
\rightarrow
\left\{
\begin{array}{l}
{\rm Tr}\,Q\,\bigl(\frac{\slash\!\!\!p+1}{2}\bigr)\,
\bigl(\frac{\gamma^0+I}{4}\bigr)\,\bigl(\frac{\slash\!\!\!p'+1}{2}\bigr)\,,\\
\sigma^{ij}\,{\rm Tr}\,Q\,\bigl(\frac{\slash\!\!\!p+1}{2}\bigr)\,
\bigl(\frac{\gamma^0+I}{2}\bigr)\,\frac{\sigma^{ij}}{4}\,
\bigl(\frac{\slash\!\!\!p'+1}{2}\bigr)\,,
\end{array}\right. 
\end{align} 
which identify the spin-independent and spin-dependent parts of the matrix
element of the arbitrary operator $Q$, respectively,
with $\sigma^{ij}$ defined by
\begin{equation}
\sigma^{ij} = \frac{i}{2}\,[\gamma^i\,,\,\gamma^j]
\stackrel{d=3}{=} \epsilon^{ijk}\,\sigma^k\,.
\end{equation}

We now perform an expansion of the integrand of the scattering amplitude
$\delta_1 H$ in Eq.~(\ref{36}) for two scales, $k\sim m$ and $k \sim m\,\alpha$.
Assuming $k\sim m$ and the external momenta $p\sim m\,\alpha$
and expanding the integrand in $\alpha$, we obtain
\begin{eqnarray}
\delta_{1} H[m] &=&\alpha^2\biggl\{ 
\sigma_1(j,q)\,\sigma_2(j,q)\,\left[-\frac{23}{36} 
+\frac{7}{12\,\epsilon}\right]
 \nonumber\\ &&  
+i\,[\sigma_1({p'_1},{p_1})+\sigma_2({p'_2},{p_2})]\,
\left[ -\frac{1}{6}- \frac{1}{4\,\epsilon}\right]
\nonumber \\ && 
+i\,[\sigma_1({p'_2},{p_2})+\sigma_2({p'_1},{p_1})]\!
\left[\frac{1}{4}\right]
 \nonumber\\ && 
  +\frac{1}{8}\,\sigma_1(j,{p_1} + {p'_1})\,
      \sigma_2(j,{p_2} + {p'_2})
 \nonumber\\ &&
  -\frac{1}{8}\,\sigma_1(j,{p_2} + {p'_2})\,
        \sigma_2(j,{p_1} + {p'_1})
 \nonumber\\ && 
+\frac{17}{72}\,\sigma_1(j,{p_1} - {p_2} + {p'_1} - {p'_2})
 \nonumber\\ && \times
\,\sigma_2(j,{p_1} - {p_2} + {p'_1} - {p'_2}) \biggr\}\,.
\end{eqnarray}
A similar expansion with the assumption that $k\sim m\,\alpha$ leads to
\begin{align}
\delta_{1} H[&m\alpha]  =\alpha^2\biggl\{ 
\sigma_1(j,q)\,\sigma_2(j,q)\,\left[-\frac{5}{12} 
-\frac{1}{4\,\epsilon}+ \frac{1}{2}\,\ln (q)\right]
 \nonumber\\ &  
+i\,[\sigma_1({p'_1},{p_1})+\sigma_2({p'_2},{p_2})]\,
\left[ \frac{7}{12}- \frac{1}{12\,\epsilon} + \frac{1}{6}\,\ln (q)\right]
\nonumber \\ & 
+i\,[\sigma_1({p'_2},{p_2})+\sigma_2({p'_1},{p_1})]\!
\left[\frac{2}{3} - \frac{2}{3\,\epsilon} + \frac{4}{3}\,\ln (q)\right]
 \biggr\}.
\end{align}
The sum of $\delta_{1} H[m]$ and $\delta_{1} H[m\alpha]$ is
\begin{eqnarray}
\delta_1 H &=& \alpha^2\biggl\{ 
\sigma_1(j,q)\,\sigma_2(j,q)\,\left[-\frac{19}{18} 
+\frac{1}{3\,\epsilon}\,+ \frac{1}{2}\,\ln (q)\right]
 \nonumber\\ &&  
+i\,[\sigma_1({p'_1},{p_1})+\sigma_2({p'_2},{p_2})]\,
\left[ \frac{5}{12}- \frac{1}{3\,\epsilon} + \frac{1}{6}\,\ln (q)\right]
\nonumber \\ && 
+i\,[\sigma_1({p'_2},{p_2})+\sigma_2({p'_1},{p_1})]\!
\left[\frac{11}{12} - \frac{2}{3\,\epsilon} + \frac{4}{3}\,\ln (q)\right]
 \nonumber\\ && 
  +\frac{1}{8}\,\sigma_1(j,{p_1} + {p'_1})\,
      \sigma_2(j,{p_2} + {p'_2})
 \nonumber\\ &&
  -\frac{1}{8}\,\sigma_1(j,{p_2} + {p'_2})\,
        \sigma_2(j,{p_1} + {p'_1})
 \nonumber\\ && 
+\frac{17}{72}\,\sigma_1(j,{p_1} - {p_2} + {p'_1} - {p'_2})
 \nonumber\\ && \times
\,\sigma_2(j,{p_1} - {p_2} + {p'_1} - {p'_2}) \biggr\},
\label{108}
\end{eqnarray} 
where $\sigma(j,q)=\sigma^{ji}\,q^i$, and $q = \sqrt{\bfq^2}$.

The third scale $k\sim m\,\alpha^2$ requires a more accurate treatment since
any number of the electron-nucleus Coulomb photon exchanges contribute to the
same order. This low energy part can be represented in the Coulomb gauge as 
\begin{align} \label{Elep}
E_{LE} &\ = e^2\,\int_0^{\infty}
\,\frac{d^dk}{(2\,\pi)^d\,2\,k}\,
\biggl(\delta^{ij}-\frac{k^i\,k^j}{k^2}\biggr)
\nonumber \\ &\times
\delta\left\langle\phi\left|p_1^i\,\frac{1}{E_0-H_0-k}\,p_2^j
\right|\phi\right\rangle + (1 \leftrightarrow 2)\,,
\end{align}
where the symbol $\delta \lbr \ldots \rbr$ 
stands for the first-order perturbation correction
of the matrix element $\lbr \ldots \rbr$ by the ($d$-dimension generalization of the) 
Breit-Pauli Hamiltonian $H^{(4)}$, which implies perturbations of the
reference-state wave function $\phi$, the energy $E_0$, and the zeroth-order
Hamiltonian $H_0$.   

The expression (\ref{Elep}) involves the Coulomb Green function, which is not
known for the arbitrary dimension. This problem is solved by splitting the
integral over $k$ into two parts, 
\begin{eqnarray} \label{split}
\int_0^\infty d k = \int_0^{\Lambda} d k + 
\int_{\Lambda}^\infty d k\,,
\end{eqnarray}
with $\Lambda = m\,(Z\,\alpha)^2\,\lambda$ and $\lambda$ being a dimensionless cutoff
parameter. 
The two corresponding parts of Eq.~(\ref{Elep}) will be referred to
as  ${\cal E}_{LE}$ and  $\delta_2 E$. 
It is assumed that in these parts the expansion is performed first in the small $\epsilon$
and next in the large $\lambda$.
The first part ${\cal E}_{LE}$ has a finite limit at $d=3$. It will be
evaluated in Sec.~\ref{sec:bethe} together with other low-energy contributions.

We now turn to the evaluation of the second part $\delta_2 E$. The
spin-dependent part of the Breit-Pauli Hamiltonian in $d$ dimensions is the sum of the
electron-electron part $H_{ee}$ and the electron-nucleus part $H_{eN}$,
\begin{eqnarray}
\delta H_{ee} &=& \frac{e^2}{4\,m^2\,q^2}\,\bigl[
i\,{p'_1}^i\,{p_1}^j(\sigma_1^{ij}+2\,\sigma_2^{ij}) 
\nonumber \\ &&
+ i\,{p'_2}^i\,{p_2}^j(\sigma_2^{ij}+2\,\sigma_1^{ij})
-\sigma_1^{ik}\,\sigma_2^{jk}\,q^i\,q^j\bigr]\,,\label{48}
\end{eqnarray}
\begin{equation}
\delta H_{eN} = -\frac{Z\,e^2}{4\,q^2}\,\bigl[
i\,\sigma_1^{ij}\,{p''_1}^i\,p_1^j +
i\,\sigma_1^{ij}\,{p''_2}^i\,p_2^j\bigr]\,.
\label{49}
\end{equation}
Only the electron-electron part $H_{ee}$ contribute to $\delta_2 E$.
Since $k$ is much larger than $H_0-E_0$,
we expand the integrand to yield
\begin{align}
\delta& \left\langle\phi\left|p_1^i\,\frac{1}{E_0-H_0-k}\,p_2^j
 \right|\phi\right\rangle + (1 \leftrightarrow 2)
 \nonumber \\
&  = \frac{1}{k^2}\,\delta\left\langle\phi\left|p_1^i\,(H_0-E_0)\,p_2^j
\right|\phi\right\rangle + (1 \leftrightarrow 2)
\nonumber \\ &=
\frac{1}{k^2}\,\delta\left\langle\phi\left|
[p_1^i,\,[H_0-E_0\,,\,p_2^j]]\right|\phi\right\rangle \nonumber \\ &=
\frac{2}{k^2}\,\left\langle\phi\left|
[p_1^i,\,[V\,,\,p_2^j]]\,\frac{1}{(E_0-H_0)'}\,\delta H_{ee}\right|\phi\right\rangle
 \nonumber \\ &+
\frac{1}{k^2}\,\left\langle\phi\left|
[p_1^i,\,[\delta H_{ee}\,,\,p_2^j]]\right|\phi\right\rangle \,.
\end{align}
The first term in the above expression is the second-order perturbation
correction, which is already included into $E_S$, Eq.~(\ref{ES}). The contribution of
the second term is 
\begin{align}
\delta_2 E = e^2\,\frac{d-1}{d}\,\int_\Lambda^{\infty}
\,\frac{d^dk}{(2\,\pi)^d\,2\,k^3}\,
\left\langle\phi\left|
[p_1^i,\,[\delta H_{ee}\,,\,p_2^i]]\right|\phi\right\rangle\,.
\end{align}
After expanding this expression in $\epsilon =(3-d)/2$ and then in $\alpha$,
the result is represented by the expectation value of the effective operator
$\delta_2 H$,
\begin{eqnarray}
\delta_2 H &=& \alpha^2\left[\frac{5}{9}+\frac{1}{3\,\epsilon}+
\frac{2}{3}\,\ln[(Z\,\alpha)^{-2}]-\frac{2}{3}\,\ln(2\,\lambda)\right]\,
\nonumber \\ && \times 
[i\,\sigma_1({p'_1},{p_1})+i\,\sigma_2({p'_2},{p_2})
+2\,i\,\sigma_1({p'_2},{p_2})
\nonumber \\ && 
+2\,i\,\sigma_2({p'_1},{p_1})
-\sigma_1(j,q)\,\sigma_2(j,q)]\,.
\end{eqnarray}
So, the contribution due to the photon exchange is given by the sum of the expectation
values of $\delta_1 H$ and $\delta_2 H$ and by the low-energy contribution
${\cal E}_{LE}$. 

When calculating expectation values of effective operators between the triplet $P$
states, some simplifications can be performed. The first one is that the 
expectation value of the Dirac $\delta$ function with both momenta
on the right (left) hand side vanishes. The second one is that
the expectation value
of $\sigma_1$ is equal to that of $\sigma_2$. As a result, the 
sum of $\delta_1 H$ and $\delta_2 H$ can be written as
\begin{align} \label{HE}
H_E &\ = \delta_1 H+\delta_2 H 
 \nonumber \\ & 
= \alpha^2\,\biggl[6+4\ln[(Z\,\alpha)^{-2}] -4\,\ln(2\,\lambda)
+3\,\ln q\biggr]i\,\sigma_1({p'_1},{p_1})
\nonumber \\ & 
+\alpha^2\,\biggl[-\frac{23}{9}
-\frac{2}{3}\,\ln[(Z\,\alpha)^{-2}] +\frac{2}{3}\,\ln(2\,\lambda)
+\frac{1}{2}\,\ln q\biggr]
\nonumber \\&\times
\sigma_1(j,q)\,\sigma_2(j,q)\,.
\end{align}


\subsection{Radiative corrections} \label{sec:rad}

The $m\,\alpha^7$ contribution induced by the radiative corrections is also split into
the high and low energy parts. We argue that the high energy part can be 
accounted for by using the electromagnetic form factors $F_1$ and $F_2$
and the Uehling  correction to the Coulomb potential $F_V$,
\begin{eqnarray}
F_1(-\bfq^{\,2}) &=&
1+\frac{\alpha}{\pi}\left(\frac{1}{8}+\frac{1}{6\,\epsilon}\right)\,\bfq^{\,2}\,,\nonumber \\
F_2(-\bfq^{\,2}) &=& \frac{\alpha}{\pi}\left(\frac{1}{2}-\frac{1}{12}\,\bfq^{\,2}\right)\,,\nonumber \\
F_V(-\bfq^{\,2}) &=&  \frac{\alpha}{\pi}\,\frac{1}{15}\,\bfq^{\,2}\,.
\end{eqnarray}
In principle, there are also corrections quadratic
in the electromagnetic field. However, one can demonstrate that such
terms formed out of $\bfE$, $\bfB$, $\bfp$, and $\bsigma$
contribute only to higher orders and can be neglected.

Corrections induced by the slope of the electromagnetic form factors
are obtained by rederiving the Breit-Pauli Hamiltonian, Eqs.~(\ref{48}) and (\ref{49}),
with employing modified electromagnetic vertices. The resulting effective operator is
\begin{eqnarray} \label{H3}
\delta_3 H &=& \pi\,Z\,\alpha(F'_1+2\,F'_2+F_V')
i\,[\sigma_1(p_1'',p_1) + \sigma_2(p_2'',p_2)]
\nonumber \\ && -
\pi\,\alpha(2\,F'_1+2\,F'_2+F_V')
i\,[\sigma_1(p_1',p_1) + \sigma_2(p_2',p_2)]
\nonumber \\ &&
-2\,\pi\,\alpha(2\,F'_1+F'_2+F_V')
i\,[\sigma_1(p_2',p_2) + \sigma_2(p_1',p_1)]
\nonumber \\ &&
+\pi\,\alpha(2\,F'_1+2\,F'_2+F_V')\,
\sigma_1(j,q)\,\sigma_1(j,q)\,,
\label{10}
\end{eqnarray}
where $p''$ is the momentum scattered off 
the Coulomb potential of the nucleus.

The low-energy part of the radiative contribution is written in a form
similar to Eq. (\ref{Elep}), 
\begin{align} \label{Rlep}
E_{LR} &\ = e^2\,\int_0^{\infty}
\,\frac{d^dk}{(2\,\pi)^d\,2\,k}\,\left(\delta^{ij}-\frac{k^i\,k^j}{k^2}\right)
 \nonumber \\ & \times
\delta\left\langle\phi\left|p_1^i\,\frac{1}{E-H-k}\,p_1^j
\right|\phi\right\rangle + (1 \rightarrow 2)\,.
\end{align}
Here, $\delta$ denotes the first-order perturbation correction due to both 
parts of the Breit-Pauli Hamiltonian, the electron-electron part 
$\delta H_{ee}$ in Eq. (\ref{48})
and the electron-nucleus part $\delta H_{eN}$ in Eq. (\ref{49}). 
Introducing the splitting parameter $\lambda$,
we separate $E_{LR}$ into two parts, ${\cal E}_{LR}$ and 
$\delta_4 E$, which correspond to the first and the second term in
Eq.~(\ref{split}), respectively. 

The evaluation of $\delta_4 E$ is similar to that of the photon
exchange part. It yields $\delta_4 E = \lbr \delta_4 H \rbr$, where the effective
Hamiltonian is
\begin{eqnarray} \label{H4}
\delta_4 H &=& \alpha^2\,\left[\frac{5}{9}+\frac{1}{3\,\epsilon}
+\frac{2}{3}\,\ln[(Z\,\alpha)^{-2}]-\frac{2}{3}\,\ln(2\,\lambda)\right]
\nonumber \\ && \times
\biggl[\frac{i\,Z}{2}\,\sigma_1(p''_1,p_1)+
\frac{i\,Z}{2}\,\sigma_2(p''_2,p_2)
\nonumber \\ &&
-i\,\sigma_1(p'_1,p_1) -i\,\sigma_2(p'_2,p_2) 
-2\,i\,\sigma_2(p'_1,p_1) 
\nonumber \\ &&
-2\,i\,\sigma_1(p'_2,p_2)
+\sigma_1(j,q)\,\sigma_2(j,q)\biggr].
\label{12}
\end{eqnarray}  

The total radiative correction is the sum of $\delta_3H$, $\delta_4H$, and the
low-energy contribution ${\cal E}_{LR}$. The sum of $\delta_3H$ and $\delta_4H$
can be simplified further by using the symmetry $1\leftrightarrow 2$,
with the result
\begin{eqnarray} \label{HR}
H_R &=& Z\,\alpha^2\left[\frac{91}{180}+\frac{2}{3}\,
\ln[(Z\,\alpha)^{-2}]-\frac{2}{3}\,\ln(2\,\lambda)\right]i\,\sigma_1(p''_1,p_1) 
\nonumber \\ &&
-\alpha^2\left[\frac{21}{10}+4\,\ln[(Z\,\alpha)^{-2}]-4\,\ln(2\,\lambda)\right]\,
i\,\sigma_1(p'_1,p_1)\,,
\nonumber \\ &&
+\alpha^2\left[\frac{73}{180}+\frac{2}{3}\,\ln[(Z\,\alpha)^{-2}]
-\frac{2}{3}\,\ln(2\,\lambda)\right]
\nonumber \\ && \times
\sigma_1(j,q)\,\sigma_2(j,q)\,.
\end{eqnarray}

\subsection{$\bm{Q}$-operators}
It is convenient to consider the sum of Eqs. (\ref{HE}) and (\ref{HR}),
$H_Q = H_E+H_R$, as several logarithmic terms cancel out. The results is
\begin{eqnarray} 
H_Q &=&Z\,\alpha^2\left[\frac{91}{180}+\frac{2}{3}\,
\ln[(Z\,\alpha)^{-2}]-\frac{2}{3}\,\ln(2\,\lambda)\right]
\nonumber \\ && \times
i\,\sigma_1(p''_1,p_1) 
+\alpha^2\left[\frac{39}{10}+3\,\ln q\right]\,i\,\sigma_1(p'_1,p_1)
\nonumber \\ &&
+\alpha^2\left[-\frac{43}{20}+\frac{\ln q}{2}\right]\,
\sigma_1(j,q)\,\sigma_2(j,q).
\end{eqnarray}

For numerical calculations, we have to obtain the coordinate space
representation of $H_Q$, which involves singular operators and requires a
proper definition. We introduce the following operators 
\begin{eqnarray}
\int \frac{d^3 q}{(2\,\pi)^3}\,e^{i\,\bfq\cdot\bfr}\,4\,\pi\,(1-\ln q) = \frac{1}{r^3}\,,
\end{eqnarray}
\begin{align}
\int \frac{d^3 q}{(2\,\pi)^3}\,e^{i\,\bfq\cdot\bfr}\,\frac{4\,\pi}{15}\,
\biggl(q^i\,q^j-\frac{\delta^{ij}}{3}\,q^2\biggr)\,& \biggl(\ln
q-\frac{23}{15}\biggr) 
\nonumber \\ &
=
\frac{1}{r^7}\,\biggl(r^i\,r^j-\frac{\delta^{ij}}{3}\,r^2\biggr)\,.
\end{align}
The coordinate-space representation of these operators is defined through
their integrals with the arbitrary function $f$ that is smooth at origin,
\begin{align} \label{def1}
\int d^3 r \frac{1}{r^3}\,f(\bfr) \equiv \lim_{\epsilon\rightarrow 0} \int
d^3 r\,&\biggl[\frac{1}{r^3}\,\theta(r-\epsilon)
\nonumber \\ &
+ 4\,\pi\,\delta^3(r)\,(\gamma+\ln\epsilon)\biggr]\,f(\bfr) \,,
\end{align}
\begin{align} \label{def2}
\int d^3 r\,\frac{1}{r^7}&\,\biggl(r^i\,r^j-\frac{\delta^{ij}}{3}\,r^2\biggr)\,f(\bfr) \equiv
\lim_{\epsilon\rightarrow 0} \int
d^3 r\,
\nonumber \\ &
\biggl[\frac{1}{r^7}\,\biggl(r^i\,r^j-\frac{\delta^{ij}}{3}\,r^2\biggr)\theta(r-\epsilon)
\nonumber \\ &
+ \frac{4\,\pi}{15}\,\delta^3(r)\,(\gamma+\ln\epsilon)\,
\biggl(\partial^i\,\partial^j-\frac{\delta^{ij}}{3}\,\partial^2\biggr)\biggr]\,f(\bfr)\,,
\end{align}
where we assume that $r$ is expressed in atomic units.

With these definitions, the coordinate-space representation of $H_Q$ is
(in atomic units)
\begin{eqnarray} \label{HQ}
H_Q &=& Z\,\alpha^7\left(\frac{91}{180} +\frac{2}{3}\,
\ln[(Z\,\alpha)^{-2}]-\frac{2}{3}\,\ln(2\,\lambda)\right)
\nonumber \\ && \times
\bigl(i\,\bfp_1\times\delta^3(r_1)\,\bfp_1\cdot\bsigma_1\bigr) \nonumber \\ &&
+\alpha^7\left(-\frac{83}{60}+\frac{\ln\alpha}{2}\right)\,
(\bsigma_1\cdot\bnabla)\,(\bsigma_2\cdot\bnabla)
\delta^3(r)\nonumber \\ &&
-\alpha^7\,\frac{15}{8\,\pi}\,
\frac{1}{r^7}\,(\bsigma_1\cdot\bfr)\,(\bsigma_2\cdot\bfr)\nonumber \\ &&
+\alpha^7\,\left(\frac{69}{10}+3\,\ln\alpha\right)\,
i\,\bfp_1\times\delta^3(r)\,\bfp_1\cdot\bsigma_1\nonumber \\ &&
-\alpha^7\,\frac{3}{4\,\pi}\,i\,\bfp_1\times \frac{1}{r^3}\,
\bfp_1\cdot\bsigma_1\,. \label{hq}
\end{eqnarray}
The above equation is written in atomic units because the definitions
(\ref{def1}) and (\ref{def2}) are formulated in this unit system.  
Other formulas in the present paper
are written in relativistic units. The $\lambda$-dependent term in
Eq.~(\ref{HQ}) cancels with the corresponding contribution in Eq.~(\ref{bethe}).
The logarithmic in $\alpha$ part of $H_Q$ agrees 
with the results of Refs. \cite{zhang:96:prl,pachucki:99:jpb}.

\subsection{Anomalous magnetic moment correction to the $\bm{m\,\alpha^6}$ operators}

The remaining contribution to $H^{(7)}_{\rm fs}$ is the amm 
correction to the spin-dependent $m\,\alpha^6$ operators. It does not lead to any
divergences and therefore can be calculated without any regularization.
The derivation of the amm part of  $H^{(7)}_{\rm fs}$ is done
with the help of the NRQED Hamiltonian obtained
by the FW transformation of the Dirac Hamiltonian with
the electron magnetic moment anomaly included.
The resulting Hamiltonian, with higher-order spin-independent terms omitted,
is \cite{jentschura:05:sese}
\begin{eqnarray}
H_{FW} &=& \frac{\bpi^2}{2}+e\,A^0 -\frac{e}{2}\,(1+a_e)\,
\bsigma\cdot\bfB -\frac{\bpi^4}{8}
\nonumber \\ &&
-\frac{e}{8}\,(1+2\,a_e)\,[\bnabla\cdot\bfE+
\bsigma\cdot(\bfE\times\bpi-\bpi\times\bfE)]
\nonumber \\ &&
+\frac{e}{8}\bigl(\{\bsigma\cdot\bfB,\bpi^2\}
+a_e\,\{\bpi\cdot\bfB, \bpi\cdot\bsigma\}\bigr)
\nonumber \\ &&
+\frac{(3+4\,a_e)}{32}\,\{\bfp^{\,2},e\,\bfE\times\bfp\cdot\bsigma\}\,.
\label{hfw}
\end{eqnarray} 
We use the opportunity to correct
the misprint in Ref.~\cite{pachucki:06:prl:he} where the last term was typed
with an incorrect prefactor. As demonstrated in Ref.~\cite{pachucki:05:hamil}, 
all spin-dependent operators to order $m\,\alpha^6$ can be obtained from $H_{FW}$. The
derivation of the amm correction to the $m\,\alpha^6$ operators is very much similar. 

We start with the general expression for the one-photon exchange amplitude
between the electron $a$ and the electron $b$,
\begin{align} \label{1phot}
& \langle \delta H\rangle  = e^2\int\frac{d^4
k}{(2\,\pi)^4\,i}\,G_{\mu\nu}(k)\,
\nonumber \\ & \times 
\biggl\{
\biggl\langle\phi\biggl|\jmath^\mu_a(k)\,e^{i\,\bfk\cdot\bfr_a} 
\frac{1}{E_0-H_0-k^0+i\,\epsilon}
\jmath^\nu_b(-k)\,e^{-i\,\bfk\cdot\bfr_b}\,\biggr|\phi\biggr\rangle 
\nonumber \\ &
+\biggl\langle\phi\biggl|\jmath^\mu_b(k)\,e^{i\,\bfk\cdot\bfr_b} 
\frac{1}{E_0-H_0-k^0+i\,\epsilon}
\jmath^\nu_a(-k)\,e^{-i\,\bfk\cdot\bfr_a}\,
\biggr|\phi\biggr\rangle \biggr\}\,,
\end{align}
where $G_{\mu\nu}$ is the photon propagator in the Coulomb gauge,
\begin{eqnarray}
G_{\mu\nu}(k) = \left\{
\begin{array}{ll}
{-\frac{1}{\bfk^2}}\,, & \mu = \nu = 0\,, \\
{\frac{-1}{k_0^2-\bfk^2 +i\,\epsilon}}\Bigl(\delta_{ij}-{\frac{{k}_i {k}_j}
{\bfk^2}\Bigr)}\,, & \mu =i, \nu =j\,,
\end{array}
\right.
\end{eqnarray}
$\phi$ is an eigenstate of $H_0$ and $\jmath^\mu_a$ is the operator of the
electromagnetic current for particle $a$. In the following, we will
consider separately the exchange by the Coulomb $G_{00}$
and the transverse $G_{ij}$ photons. The expression for the electromagnetic
current  $\jmath^\mu$ is obtained from the Hamiltonian $H_{FW}$ as a
coefficient that multiplies the electromagnetic potential $A_{\mu}$.
The first terms of the nonrelativistic expansion of the current are
\begin{equation}
\jmath^0(\bfk) = 1 +\frac{i}{4\,m}\,\bsigma\cdot\bfk\times\bfp - \frac{1}{8\,m^2}\bfk^{\,2}+\ldots\,,
\end{equation}
for the $\jmath^0$ component
and 
\begin{equation}
\vec{\jmath}(\bfk) = \frac{\bfp}{m} +
\frac{i}{2\,m}\,\bsigma\times\bfk+\ldots\,,
\end{equation}
for the $\vec{\jmath}$ component.

The main part of the calculation is performed in the nonretardation approximation,
which consists in setting $k^0=0$ in the photon propagator $G_{\mu\nu}(k)$ and
in the current $\jmath(k)$;
the retardation corrections are considered separately.
Employing the nonretardation approximation and the symmetry
$k^0 \leftrightarrow -k^0$, 
the integration over $k^0$ is carried out as
\begin{equation}
\frac{1}{2} \int\frac{d\,k^0}{2\,\pi\,i}\,
\biggl[\frac{1}{-\Delta E-k^0+i\,\epsilon}+
\frac{1}{-\Delta E+k^0+i\,\epsilon}\biggr] = -\frac{1}{2}\,.
\end{equation}
The one-photon exchange amplitude in the nonretardation approximation thus is
\begin{align} \label{1phot1}
\langle\phi|\delta H|\phi\rangle &\ = -e^2\int\frac{d^3
k}{(2\,\pi)^3}\,G_{\mu\nu}(\bfk)\,
\nonumber \\ & \times
\biggl\langle\phi\biggl|\jmath^\mu_a(\bfk)\,e^{i\,\bfk
    \cdot(\bfr_a-\bfr_b)} \,\jmath^\nu_b(-\bfk)\,\biggr|\phi\biggr\rangle\,.
\end{align}
To the leading order, the current does not depend on $\bfk$ and the $\bfk$ integration 
gives the coordinate-space representation of the photon propagator 
in the nonretardation approximation,
\begin{align}
G_{\mu\nu}(\bfr) &\ = \int \frac{d^3 k}{(2\,\pi)^3}\, e^{i\,\bfk\cdot\bfr}\, G_{\mu\nu}(\bfk) 
\nonumber \\ &
= \frac{1}{4\,\pi}\,\left\{
\begin{array}{ll}
-\frac{1}{r}\,,  & \mu = \nu = 0\,, \\
\frac{1}{2\,r}\Bigl(\delta_{ij}+{\frac{{r}_i {r}_j}
{r^{\,2}}\Bigr)}\,,& \mu =i, \nu =j\,.
 \end{array}
\right.
\end{align}
One easily recognizes that in the nonrelativistic limit
$G_{00}$ is the Coulomb interaction. This term is already included
in $H_0$, which means that the nonrelativistic Coulomb interaction
has to be excluded from the perturbative expansion. Next-order terms
resulting from the expansion of $\jmath^0$ and $\vec{\jmath}$ lead to the Breit-Pauli 
Hamiltonian. 

We are interested in the expansion terms that yield effective
operators of order $m\,\alpha^6\,a_e$. Their derivation is analogous to
that of the $m\,\alpha^6$ Hamiltonian in Ref.~\cite{pachucki:05:hamil}, 
the only difference being that the corresponding 
amm prefactors should be retained. These
prefactors will give us the required effective operator, which will be
denoted as $H^{(6)}_{a_e}$. It is expressed as a sum of various contributions 
\begin{equation}
H^{(6)}_{a_e} =\sum_{i=1}^{8} \delta H_i\,,\label{72}
\end{equation}
which are calculated in the following. 

$\delta H_{1}$ is the correction due to the last term
in $H_{FW}$ in Eq. (\ref{hfw}). This term involves only $A^0$ and its
gradient,  so the nonretardation approximation 
is valid here. $\delta H_{1}$ includes the Coulomb interaction
between the electron and the nucleus and between the electrons.
We denote by  $V$ the nonrelativistic interaction potential 
\begin{equation}
V \equiv -\sum_a \frac{Z\,\alpha}{r_a} + \sum_{a>b}\,\sum_b\frac{\alpha}{r_{ab}}\,,
\label{40}
\end{equation}
and by ${\cal E}_a$ the static electric field at the position of particle $a$
\begin{equation}
e\,\vec{\cal E}_a \equiv -\nabla_a V =  
-Z\,\alpha\,\frac{\bfr_a}{r_a^3} +\sum_{b\neq a}\alpha\,\frac{\bfr_{ab}}{r_{ab}^3}\,,
\label{41}
\end{equation}
and write $\delta H_{1}$  as
\begin{eqnarray}
\delta H_{1} &=& \sum_a
\frac{3+4\,a_{e}}{32\,m^4}\,\bsigma_a\cdot
\Bigl(p_a^2\,e\,\vec{\cal E}_a\times\bfp_a +
                       e\,\vec{\cal E}_a\times\bfp_a\,p_a^2\Bigr) \label{42}
\nonumber \\ &\stackrel{a_e}{=}&
a_e\,\biggl(-\frac{Z\,\alpha}{2}\,p_1^2\,\frac{\bfr_1}{r_1^3}\times\bfp_1\cdot\bsigma_1
+\frac{\alpha}{2}\,p_1^2\,\frac{\bfr}{r^3}\times\bfp_1\cdot\bsigma_1\biggr)\,,
\nonumber \\ 
\end{eqnarray}
where by $\stackrel{a_e}{=}$ we denote that the equation is valid 
modulo terms independent on  $a_e$.

$\delta H_{2}$ is the correction 
to the Coulomb interaction between electrons
which comes from the 5$^{\rm th}$ term in $H_{FW}$, namely
\begin{equation}
-\frac{e}{8}\,(1+2\,a_e)\,\Bigl[\vec{\nabla}\cdot\bfE + \bsigma\cdot
\bigl(\bfE\times\bfp-\bfp\times\bfE \bigr)\Bigr]\,.
\label{43}
\end{equation}
If the interaction of both electrons is modified by this term,
the nonretardation approximation holds and Eq.~(\ref{1phot1}) yields
\begin{widetext}
\begin{eqnarray}
\delta H_{2} &=& \sum_{a>b}\sum_b
\int d^3 k\,\frac{e^2}{k^2}\,\frac{(1+2\,a_e)^2}{64}\,
\biggl(k^2 +2\,i\,\bsigma_a\cdot\bfp_a\times\bfk\biggr)\,
e^{i\,\bfk\cdot\bfr_{ab}}\,
\biggl(k^2 +2\,i\,\bsigma_b\cdot\bfk\times\bfp_b\biggr)
\nonumber \\ &\stackrel{\rm fs}{=}& 
e^2\,\frac{(1+2\,a_e)^2}{16}\,\int d^3 k\,
\biggl(i\,\bsigma_1\cdot\bfp_1\times\bfk\,e^{i\,\bfk\cdot\bfr}
-\bsigma_1\cdot\bfp_1\times\bfk\,\frac{e^{i\,\bfk\cdot\bfr}}{k^2}\,\bsigma_2\cdot\bfk\times\bfp_2\biggr)
 \label{44}\\ &\stackrel{a_e}{=}&
a_e\biggl[
\frac{3\,i}{2}\,\frac{\alpha}{r^5}\,\bfr\times(\bfr\cdot\bfp_2)\,\bfp_1\cdot\bsigma_1
-\frac{3}{4}\,\frac{\alpha}{r^5}\,\bfr\times(\bfr\times\bfp_1\cdot\bsigma_1)\,\bfp_2\cdot\bsigma_2
-\frac{\alpha}{4\,r^3}\,(\bfp_1\cdot\bsigma_2)\,(\bfp_2\cdot\bsigma_1)\biggr]\,,
\nonumber
\end{eqnarray}
\end{widetext}
where by $\stackrel{\rm fs}{=}$ we denote the equation that is valid
modulo spin independent terms. To make the comparison with previous
calculations more transparent, we transformed operators 
to the same form as in the original DK derivation.

$\delta H_3$ is the relativistic correction 
to the transverse photon exchange. 
The first electron is coupled to $\bfA$ by the nonrelativistic term,
\begin{equation}
-\frac{e}{m}\,\bfp\cdot\bfA -\frac{e}{2\,m}\,\bsigma\cdot\bfB\,,
\end{equation} 
and the second one, by the relativistic correction, i.e., the last but one term in Eq. (\ref{hfw})\,,
\begin{align}
\frac{e}{8}\bigl(\{\bsigma\cdot\bfB,\bpi^2\}
&+a_e\,\{\bpi\cdot\bfB, \bpi\cdot\bsigma\}\bigr) \rightarrow
\frac{e}{8}\,\bigl[p^2\,2\,\bfp\cdot\bfA
\nonumber \\ &
+2\,\bfp\cdot\bfA\,p^2+ 
\bsigma\cdot\bfB\,p^2+p^2\,\bsigma\cdot\bfB
\nonumber \\ &
+a_e\bigl(\bfp\cdot\bfB\,\bfp\cdot\bsigma+
\bfp\cdot\bsigma\,\bfp\cdot\bfB\bigr)\bigr]\,,
\end{align}
It is sufficient to calculate $\delta H_3$ in the nonretardation
approximation, in which any correction
can be simply obtained by replacing the magnetic field $\bfA$ by 
the static field $\vec {\cal A}_{a}$,
\begin{align}
e\,{\cal A}^i_{a} \equiv \sum_{b\neq a} &\  \left[ \frac{\alpha}{2\,r_{ab}}
\biggl(\delta^{ij}+\frac{r_{ab}^i\,r_{ab}^j}{r_{ab}^2}\biggr)\,p_b^j
\right.
\nonumber \\ & 
\left. +
\frac{\alpha\,(1+a_e)}{2}\frac{\bigl(\bsigma_b\times\bfr_{ab}\bigr)^i}{r_{ab}^3}
\right]\,.\label{52}
\end{align}
The result then is 
\begin{widetext}
\begin{eqnarray}
\delta H_3 &=&
\sum_a\,\frac{e}{8}\,
\Bigl[2\,p_a^2\,\bfp_a\cdot\vec {\cal A}_{a} 
+2\,\bfp_a\cdot\vec {\cal A}_{a}\,p_a^2
+ p_a^2\,\bsigma_a\cdot\bnabla_a\times\vec {\cal A}_{a} 
+ \bsigma_a\cdot\bnabla_a\times\vec {\cal A}_{a}\,p_a^2
+a_e\bigl(\bfp_a\cdot(\bnabla_a\times \vec{\cal A}_a)\,\bfp_a\cdot\bsigma_a 
\nonumber \\ &&
+ \bfp_a\cdot\bsigma_a\,\bfp_a\cdot(\bnabla_a\times \vec{\cal A}_a)\bigr)
\Bigr]\label{53}\\ &\stackrel{DK}{=}&
\frac{e}{2}\bigl[2\,p_1^2\,\bfp_1\cdot\vec{\cal A}_1 + p_1^2\,\bnabla_1\times\vec
 {\cal A}_1\cdot\bsigma_1 + a_e\,\bfp_1\cdot\bsigma_1\,\bfp_1\cdot\bnabla_1\times\vec{\cal A}_1\bigr]
\nonumber \\ &\stackrel{a_e}{=}&
a_e\,\alpha\,\biggl[\frac{1}{2}\,p_1^2\,\frac{\bfr}{r^3}\times\bfp_1\cdot\bsigma_1
+\frac{3}{4}\,p_1^2\,\frac{\bfr\cdot\bsigma_1\;\bfr\cdot\bsigma_2}{r^5}
-\frac{1}{2}\,\bfp_1\cdot\bsigma_1\,\bfp_1\times\frac{\bfr}{r^3}\cdot\bfp_2
-\frac{1}{4}\,\bfp_1\cdot\bsigma_1\;\bfp_1\cdot\bsigma_2\,\frac{1}{r^3}
+\frac{3}{4}\,\bfp_1\cdot\bsigma_1\,\bfp_1\cdot\bfr\,\frac{\bfr}{r^5}\cdot\bsigma_2\biggr]\,.
\nonumber\\
\end{eqnarray}
\end{widetext}
where by $\stackrel{DK}{=}$ we denote the equation
which is valid on the level of the expectation value of the operator on the
triplet-state wave functions.
More explicitly, following Douglas and Kroll, we use the symmetry $(1\leftrightarrow 2)$
of the wave function to replace terms involving $\sigma_2$ by terms with $\sigma_1$.

The effective operator $\delta H_4$ originates from the coupling
\begin{equation}
\frac{e^2\,(1+2\,a_e)}{8}\,\bsigma\cdot(\bfE\times\bfA-\bfA\times\bfE)\,,
\label{54}
\end{equation}
present in the fifth term in Eq. (\ref{hfw}). 
The resulting correction is obtained by replacing the fields $\bfE$ and
$\bfA$ by the static fields produced by the other electrons, with the result
\begin{eqnarray}
\delta H_4 &=& \sum_a \frac{e^2\,(1+2\,a_e)}{8}\,\bsigma_a\cdot
\Bigl[\vec{\cal E}_a\times \vec {\cal A}_{a} - \vec {\cal
    A}_{a}\times\vec{\cal E}_a\Bigr]\label{55}\nonumber \\ 
&\stackrel{DK, a_e}{=}&
a_e\,\biggl(
\frac{3}{4}\,\frac{Z\,\alpha^2}{2\,r_1^3\,r^3}\,\bsigma_1\cdot\bfr\;\bsigma_2\cdot\bfr_1
-\frac{3}{4}\,\frac{\alpha^2}{r^6}\,\bsigma_1\cdot\bfr\;\bsigma_2\cdot\bfr
\nonumber \\ &&
-\frac{Z\,\alpha^2}{2\,r_1^3\,r}\,\bfr_1\times\bfp_2\cdot\bsigma_1
+\frac{Z\,\alpha^2}{2\,r_1^3\,r^3}\,\bfr\times\bfr_1\cdot\bsigma_1\,\bfr\cdot\bfp_2
\nonumber \\ &&
+\frac{\alpha^2}{2\,r^4}\,\bfr\times\bfp_2\cdot\bsigma_1\biggr)\,.
\end{eqnarray}

The effective operator $\delta H_5$ comes from the coupling
\begin{equation}
\frac{e^2}{2}\,\bfA^2\,,\label{56}
\end{equation}
present in the first term of Eq. (\ref{hfw}). Again, in the nonretardation
approximation, the field $\bfA_a$ can be replaced by the static field produced
by the other electrons,
\begin{equation}
\delta H_5 = \sum_a\frac{e^2}{2}\,\vec {\cal A}_a^2 
\stackrel{DK}{=} e^2\,{\cal A}_1^2 \stackrel{a_e}{=}-a_e\,\frac{\alpha^2}{2\,r^4}\,\bfr
\times\bfp_1\cdot\bsigma_1\,.
\label{57}
\end{equation}

The effective operators
$\delta H_6$ and $\delta H_7$ represent the single- and the double-spin part of the
retardation correction to the nonrelativistic single
transverse photon exchange. To calculate them, we have to return to the 
general expression for the one-photon exchange amplitude, 
Eq. (\ref{1phot}), and take the transverse part of the photon propagator,
\begin{align}
\delta E &\ = -e^2\,\int\frac{d^4k}{(2\,\pi)^4\,i}\,
\frac{1}{(k^0)^2-\bfk^2+i\,\epsilon}\,\biggl(\delta^{ij}-\frac{k^i\,k^j}
{\bfk^2}\biggr)\,
\nonumber \\ & \times
\biggl\langle\phi\biggl|\jmath^i_a(k)\,e^{i\,\bfk\cdot\bfr_a} \,
\frac{1}{E_0-H_0-k^0+i\,\epsilon}
\,\jmath^j_b(-k)\,e^{-i\,\bfk\cdot\bfr_b}\,
\biggr|\phi\biggr\rangle 
\nonumber \\ & 
+(a\leftrightarrow b)\,.\label{58}
\end{align}
We assume that the product $ \jmath^i_a(k)\;
\jmath^j_b(-k)$ contains at most a single power of $k^0$.
This allows one to perform the $k^0$ integration by encircling  
the only pole $k^0 = |\bfk|$ on the ${\rm Re}(k^0)>0$ complex half-plane
and obtain
\begin{eqnarray}
\delta E &&= e^2\,\int\frac{d^3k}{(2\,\pi)^3\,2\,k}\,
\biggl(\delta^{ij}-\frac{k^i\,k^j}{k^2}\biggr)\,
\nonumber \\ && \times
\biggl\langle\phi\biggl|\jmath^i_a(k)\,e^{i\,\bfk\cdot\bfr_a} \,
\frac{1}{E_0-H_0-k}
\,\jmath^j_b(-k)\,e^{-i\,\bfk\cdot\bfr_b}\,\biggr|\phi\biggr\rangle 
\nonumber \\ && 
+(a\leftrightarrow b)\,,\label{59}
\end{eqnarray}
where $k = |\bfk|$. The retardation expansion of the electron propagator yields
\begin{equation} 
\frac{1}{E_0-H_0-k} = -\frac{1}{k}+\frac{H_0-E_0}{k^2}
-\frac{(H_0-E_0)^2}{k^3}+\ldots\,.\label{60}
\end{equation}
The first term here contributes to the Breit-Pauli Hamiltonian and 
the second term, to $E^{(5)}$. Taking the current $\jmath^i$ in the
nonrelativistic form, the third expansion term is 
\begin{eqnarray}
\delta E &=& \sum_{a \neq b}\sum_b (-e^2)\,\int\frac{d^3k}{(2\,\pi)^3\,2\,k^4}\,
\biggl(\delta^{ij}-\frac{k^i\,k^j}{k^2}\biggr)\,
\nonumber \\ && \times 
\biggl\langle
\biggl(\bfp_a+\frac{1+a_e}{2}\,
\bsigma_a\times\bnabla_a\biggr)^i
\,e^{i\,\bfk\cdot\bfr_a}\,
(H_0-E_0)^2
\nonumber \\ && \times
\,\biggl(\bfp_b+\frac{1+a_e}{2}\,
\bsigma_b\times\bnabla_b\biggr)^j
\,e^{-i\,\bfk\cdot\bfr_b}\,\biggr\rangle\,.\label{61}
\end{eqnarray}
This is the most complicated term among the amm corrections, so
we describe its evaluation in detail. The correction is split into the double spin part 
$\delta E_6$ and the single spin part $\delta E_7$,
\begin{equation}
\delta E \stackrel{\rm fs}{=} \delta E_{6}+\delta E_{7}\,.\label{62}
\end{equation}
The double spin part is
\begin{eqnarray}
\delta E_6 &&= \sum_{a\neq b} \sum_{b}\left(-\frac{e^2}{8}\right)
\,(1+a_e)^2\,\int\frac{d^3k}{(2\,\pi)^3}
\nonumber \\ && \times
\frac{(\bsigma_a\times\bfk)\cdot(\bsigma_b\times\bfk)}{k^4}\,
\Bigl\langle
\,e^{i\,\bfk\cdot\bfr_a}\,(H_0-E_0)^2
\,e^{-i\,\bfk\cdot\bfr_b}\Bigr\rangle\,. \nonumber \\\label{63}
\end{eqnarray}
We use the commutation identity
\begin{align}
\Bigl\langle\,e^{i\,\bfk\cdot\bfr_a}\,&(H_0-E_0)^2\,e^{-i\,\bfk\cdot\bfr_b}\Bigr\rangle 
+(a\leftrightarrow b)
\nonumber \\ &
= 
\Bigl\langle\Bigl[e^{i\,\bfk\cdot\bfr_a},\Bigl[(H_0-E_0)^2, e^{-i\,\bfk
\cdot\bfr_b}\Bigr]\Bigr]\Bigr\rangle
\nonumber \\ &= 
-\frac{1}{2}\,\Bigl\langle
\bigl[p_a^2,\bigl[p_b^2,e^{i\,\bfk\cdot\bfr_{ab}}\bigr]\bigr]
\Bigr\rangle\label{64}
\end{align}
to express this correction as the expectation value of the effective operator $\delta H_{6}$,
\begin{eqnarray}
\delta H_6 &\stackrel{\rm fs}{=}& \frac{\alpha\,(1+a_e)^2}{32}\,
\biggl[p_1^2,\biggl[p_2^2\,,\,
\sigma_1^i\,\sigma_2^j\,\frac{r^i\,r^j}{r^3}
\biggr]\biggr]
\nonumber \label{65}\\ &\stackrel{a_e}{=}&
a_e\,\biggl[
\frac{3\,\alpha}{4}\,p_1^2\,\frac{\bsigma_1\cdot\bfr\,\bsigma_2\cdot\bfr}{r^5}
+\frac{\alpha}{4}\,p_1^2\,\frac{i}{r^3}\,\biggl(
(\bfr\cdot\bsigma_2)\,\bsigma_1 
\nonumber \\ &&
+(\bfr\cdot\bsigma_1)\,\bsigma_2
-\frac{3\,(\bsigma_1\cdot\bfr)\,(\bsigma_2\cdot\bfr)}{r^2}\,\bfr\biggr)\cdot\bfp_2\biggr]\,.
\nonumber \\
\end{eqnarray}

The single spin part is
\begin{eqnarray}
\delta E_7 &=& \sum_{a\neq b}\sum_b\left(-\frac{i\,e^2}{4}\right)\,(1+a_e)
\int\frac{d^3 k}{(2\,\pi)^3\,k^4}
\nonumber \\ && \times
\Bigl\langle e^{i\,\bfk\cdot\bfr_a}\,
(H_0-E_0)^2\,e^{-i\,\bfk\cdot\bfr_b}\,\bsigma_a\times\bfk\cdot\bfp_b
\nonumber \\ && 
-\bfp_a\cdot\bsigma_b\times\bfk\,e^{i\,\bfk\cdot\bfr_a}\,
(H_0-E_0)^2\,e^{-i\,\bfk\cdot\bfr_b}\Bigr\rangle\,.\label{70}
\end{eqnarray}
With the help of the integral formula
\begin{equation}
\int d^3k\,\frac{4\,\pi\,\bfk}{k^4}\,
e^{i\,\bfk\cdot\bfr} = \frac{i}{2}\,\frac{\bfr}{r}\,,\label{71}
\end{equation}
one obtains
\begin{widetext}
\begin{eqnarray}
\delta H_7 &=& \sum_{a>b}\sum_b\frac{\alpha\,(1+a_e)}{4}\biggl\{
\biggl[\bsigma_a\times\frac{\vec
    r_{ab}}{r_{ab}},\frac{p_a^2}{2}\biggr]\,\cdot
    \bigl[V,\bfp_b] + \biggl[\frac{p_b^2}{2},
\biggl[\bsigma_a\times\frac{\vec
    r_{ab}}{r_{ab}},\frac{p_a^2}{2}\biggr]\biggr]\,\cdot\bfp_b 
\nonumber \\ &&
-\bigl[\bfp_a,V\bigr]\,\cdot\biggl[p_b^2,  \bsigma_b\times\frac{\vec
    r_{ab}}{r_{ab}}\biggr]- \bfp_a\,\cdot\biggl[\frac{p_a^2}{2},
\biggl[\bsigma_a\times\frac{\vec
    r_{ab}}{r_{ab}}\,,\frac{p_b^2}{2}\biggr]\biggr]\biggr\}\nonumber
\\ &\stackrel{DK, a_e}{=}&
a_e\,\biggl(\frac{\alpha^2}{2}\,\bsigma_1\cdot\frac{\bfr}{r^4}\times\bfp_1
+\frac{Z\,\alpha^2}{2\,r}\,\bsigma_1\cdot\frac{\bfr_1}{r_1^3}\times\bfp_2
+\frac{Z\,\alpha^2}{2}\,\bsigma_1\cdot\frac{\bfr}{r^3}\times\frac{\vec
  r_1}{r_1^3}\,(\bfr\cdot\bfp_2)\nonumber \\ &&
+\frac{\alpha}{2}\,p_1^2\,\frac{i}{r}\,\bsigma_1\cdot\bfp_2\times\bfp_1
-\frac{\alpha}{2}\,p_1^2\,\frac{i}{r^3}\,(\bfr\cdot\bfp_2)\,(\bfr\times\bfp_1)\cdot\bsigma_1\biggr)\,.
\end{eqnarray}
\end{widetext}

The effective operator 
$\delta H_8$ represents the retardation correction to the single transverse
photon exchange contribution, in which one vertex is nonrelativistic, Eq.~(\ref{49}),
whereas the second comes from the fifth term in Eq.~(\ref{hfw}),
\begin{equation}
-\frac{e\,(1+2\,a_e)}{8}\,\bsigma\cdot\bigl(\bfE\times\bfp-\bfp\times\vec
 E\bigr)\,.\label{74}
\end{equation}
With the help of Eq. (\ref{59}), one obtains 
\begin{eqnarray}
\delta E_8 &=& \sum_{a\neq b}\sum_b e^2\,(1+2\,a_e)\int\frac{d^3k}{(2\,\pi)^3}\,
\biggl(\delta^{ij}-\frac{k^i\,k^j}{k^2}\biggr)\,
\nonumber \\ && \times
\frac{i}{16}\,
\biggl\langle\Bigl(e^{i\,\bfk\cdot\bfr_a}\,\bfp_a\times\bsigma_a
+\bfp_a\times\bsigma_a\,e^{i\,\bfk\cdot\bfr_a}\Bigr)^i
\nonumber \\ &&\times 
\frac{1}{E_0-H_0-k}\,\biggl(\bfp_b -
\frac{i}{2}\,\bsigma_b\times\bfk\biggr)^j\,e^{-i\,\bfk\cdot\bfr_b}\biggr\rangle
+{\rm h.c.}\,.\nonumber \\ \label{75}
\end{eqnarray}
In the expansion of $1/(E_0-H_0-k)$ in Eq. (\ref{60}) the first term vanishes
because of the Hermitian conjugation and the second term contributes to order $m\,\alpha^6\,a_e$.
After commuting $(H_0-E_0)$ on the left, one obtains the effective operator $\delta H_8$,
\begin{widetext}
\begin{eqnarray}
\delta H_8 
&=&(1+2\,a_e)\,\sum_a \biggl\{\frac{e^2}{8}\,\bsigma_a\cdot
\bigl(\vec{\cal E}_a\times \vec {\cal A}_{a} - 
\vec {\cal A}_{a}\times\vec{\cal E}_a\bigr)
+\frac{i\,e}{16}\,\biggl[
 \vec {\cal A}_{a}\cdot\bfp_a\times\bsigma_a +
\bfp_a\times\bsigma_a\cdot \vec {\cal A}_{a}\,, p_a^2\biggr]\biggr\}
\nonumber \\ &\stackrel{DK, a_e}{=}&
\delta H_5 +a_e\,\biggl[-\frac{i\,\alpha}{2}\,p_1^2\,\frac{1}{r}\,\bfp_2\times\bfp_1\cdot\bsigma_1
-\frac{i\,\alpha}{2}\,p_1^2\,\frac{1}{r^3}\,(\bfr\cdot\bfp_2)\,(\bfr\times\bfp_1)\cdot\bsigma_1
+\frac{9\,\alpha}{8}\,p_1^2\,\frac{(\bsigma_1\cdot\vec
  r)\,(\bsigma_2\cdot \bfr)}{r^5}
\nonumber \\ &&
-\frac{3\,i\,\alpha}{4}\,p_1^2\,\biggl(\frac{\bfr}{r^3}
\cdot\bsigma_1\biggr)\,(\bfp_1\cdot\bsigma_2)
-\frac{\alpha}{2}\,p_1^2\,\frac{\bfr}{r^3}\times\bfp_2\cdot\bsigma_1
-\frac{\alpha}{2}\,p_1^2\,\frac{\bfr}{r^3}\times\bfp_1\cdot\bsigma_1\biggr]\,.
\label{76}
\end{eqnarray}
\end{widetext}

Finally, after some rearrangement, the total amm correction 
can be expressed as the expectation value of 17 operators,
\begin{eqnarray} \label{EH}
E_H = \langle H^{(6)}_{a_e}\rangle = \sum_{i=1}^{17} \langle {\cal H}_i\rangle\,,
\end{eqnarray}
where
\begin{eqnarray} \label{amm0}
{\cal H}_1 &=& -\frac{Z}{4}\,p_1^2\,\frac{\bfr_1}{r_1^3}\times\bfp_1\cdot\bsigma_1\,, 
\end{eqnarray}
\begin{eqnarray} 
{\cal H}_2 &=& -\frac{3\,Z}{4}\,\frac{\bfr_1}{r_1^3}\times\frac{\bfr}{r^3}\cdot\bsigma_1\,(\bfr\cdot\bfp_2)\,, 
\end{eqnarray}
\begin{eqnarray} 
{\cal H}_3 &=& \frac{3\,Z}{4}\,\frac{\bfr}{r^3}\cdot\bsigma_1\,\frac{\bfr_1}{r_1^3}\cdot\bsigma_2\,,
\end{eqnarray}
\begin{eqnarray} 
{\cal H}_4 &=& \frac{1}{2\,r^4}\,\bfr\times\bfp_2\cdot\bsigma_1\,,
\end{eqnarray}
\begin{eqnarray} 
{\cal H}_5 &=& -\frac{3}{4\,r^6}\,\bfr\cdot\bsigma_1\,\bfr\cdot\bsigma_2\,,
\end{eqnarray}
\begin{eqnarray} 
{\cal H}_6 &=& \frac{1}{4}\,p_1^2\,\frac{\bfr}{r^3}\times\bfp_1\cdot\bsigma_1\,,
\end{eqnarray}
\begin{eqnarray} 
{\cal H}_7 &=& -\frac{1}{4}\,p_1^2\,\frac{\bfr}{r^3}\times\bfp_2\cdot\bsigma_1\,,
\end{eqnarray}
\begin{eqnarray} 
{\cal H}_8 &=& -\frac{Z}{4\,r}\,\frac{\bfr_1}{r_1^3}\times\bfp_2\cdot\bsigma_1\,,
\end{eqnarray}
\begin{eqnarray} 
{\cal H}_9 &=& -\frac{i}{2}\,p_1^2\,\frac{1}{r^3}\,\bfr\cdot\bfp_2\,\bfr\times\bfp_1\cdot\bsigma_1\,,
\end{eqnarray}
\begin{eqnarray} 
{\cal H}_{10} &=& \frac{3\,i}{4\,r^5}\,\bfr\times(\bfr\cdot \bfp_2)\,\bfp_1\cdot \bsigma_1\,,
\end{eqnarray}
\begin{eqnarray} 
{\cal H}_{11} &=& -\frac{3}{8\,r^5}\,\bfr\times(\bfr\times \bfp_1\cdot\bsigma_1)\,\bfp_2\cdot \bsigma_2\,,
\end{eqnarray}
\begin{eqnarray} 
{\cal H}_{12} &=& -\frac{1}{8\,r^3}\,\bfp_1\cdot\bsigma_2\,\bfp_2\cdot\bsigma_1\,,
\end{eqnarray}
\begin{eqnarray}
{\cal H}_{13} &=& \frac{21}{16}\,p_1^2\,\frac{1}{r^5}\,\bfr\cdot\bsigma_1\,\bfr\cdot\bsigma_2\,,
\end{eqnarray}
\begin{eqnarray} 
{\cal H}_{14} &=& -\frac{3\,i}{8}\,p_1^2\,\frac{\bfr}{r^3}\cdot\bsigma_1\,\bfp_1\cdot\bsigma_2\,,
\end{eqnarray}
\begin{eqnarray} 
{\cal H}_{15} &=& \frac{i}{8}\,p_1^2\,\frac{1}{r^3}\,\bigl(\bfr\cdot\bsigma_2\,\bfp_2\cdot\bsigma_1
         +(\bfr\cdot\bsigma_1)\,(\bfp_2\cdot\bsigma_2) 
\nonumber \\ && 
         -\frac{3}{r^2}\,\bfr\cdot\bsigma_1\,\vec
          r\cdot\bsigma_2\,\bfr\cdot\bfp_2\bigr)\,,
\end{eqnarray}
\begin{eqnarray} 
{\cal H}_{16} &=& -\frac{1}{4}\,\bfp_1\cdot\bsigma_1\,\bfp_1\times\frac{\bfr}{r^3}\cdot\bfp_2\,,
\end{eqnarray}
\begin{eqnarray} 
{\cal H}_{17} &=& \frac{1}{8}\,\bfp_1\cdot\bsigma_1\,\bigl(-\bfp_1\cdot\bsigma_2\,\frac{1}{r^3}
          +3\bfp_1\cdot\bfr\,\frac{\bfr}{r^5}\cdot\bsigma_2\bigr)\,.
 \label{amm1}
\end{eqnarray}
The operators above are intentionally written in a form very similar 
to that for the DK operators given by
Eqs.~(\ref{DK0})-(\ref{DK1}). For most of ${\cal H}_i$, there is
an one-to-one correspondence with the DK operators, all the difference
being the overall prefactors. There are only three exceptions. 
The first one is that our operator ${\cal H}_8$  cancels out in the DK
calculation, while DK operator ${\cal H}_8$ cancels out in our calculations.
The other two are related to the different 
spin structure of the next to last term in Eq. (\ref{hfw}), which leads to the
operators ${\cal H}_{16}$
and ${\cal H}_{17}$.

\subsection{Low-energy contribution}
\label{sec:bethe}

The low-energy part $E_L$ comes from the photon momenta region $k<\Lambda$,
$\Lambda$ being the cutoff parameter introduced by Eq.~(\ref{split}).
$E_L$ is a sum of the low-energy contributions due to the photon
exchange (${\cal E}_{LE}$ in Section~\ref{sec:exchange}), due to 
the radiative corrections (${\cal E}_{LR}$ in Section~\ref{sec:rad}),
and an additional  uv-finite contribution that was not considered in
previous sections.  $E_L$ can be conveniently derived from
the low energy form of the electromagnetic interaction Hamiltonian \cite{pachucki:04:lwqed},
\begin{equation}
H_I = \sum_a \biggl(-e\,\bfr_a\cdot\bfE
-\frac{e}{2\,m}\,\sigma_a^i\,r_a^j\,B^i_{,j}
-\frac{e}{4\,m^2}\,\bsigma_a\cdot\bfE_a\times\bfp_a\biggr)\,.
\label{low}
\end{equation}
This choice of the starting point for the derivation is more convenient than
the general nonrelativistic Hamiltonian in Eq. (\ref{b09}) since it makes
transparent the high degree of cancellation between various terms.
Specifically, the contributions
of the second and the third terms in $H_I$ cancel each other
and only the first term contributes to $E_L$,
\begin{eqnarray}
E_L &&= \frac{2}{3}\,\alpha\,\int_0^\Lambda \frac{d^3k}{(2\,\pi)^3\,2\,k}\,k^2\,
\nonumber \\ && \times
\delta\biggl\langle\phi\biggl|(\bfr_1 + \bfr_2)
\,\frac{1}{E-H-k}\,(\bfr_1 + \bfr_2)\biggr|\phi\biggr\rangle,
\end{eqnarray}  
where $\delta\langle\ldots\rangle$ denotes the correction to the matrix element
$\langle\ldots\rangle$ due to the Breit-Pauli Hamiltonian $H^{(4)}_{\rm fs}$.
Using the relation
\begin{equation}
\bigl[H,\bfr_1+\bfr_2\bigr] = -\frac{i}{m}(\bfp_1+\bfp_2) +
\frac{1}{4\,m^2}\,\bigl[\bfp_1\times\bsigma_1+
\bfp_2\times\bsigma_2, H_0-E_0 \bigr],
\end{equation}
$E_L$ can be transformed to the following compact form,
\begin{widetext}
\begin{eqnarray}
E_{L} &=&-\frac{2\,\alpha}{3\,\pi}\,
\delta\,\biggl\langle \phi\biggl|\,(\bfp_1+\bfp_2)\,(H-E)
\ln\left[\frac{2(H-E)}{(Z\,\alpha)^2}\right]
(\bfp_1+\bfp_2)\,\biggr|\phi\biggr\rangle
+\ln(2\,\lambda)\,\frac{2}{3}\,Z\,\alpha^2\,
\langle i\,\bfp_1\times\delta^{3}(r_1)\,\bfp_1\cdot\bsigma_1\rangle
\nonumber \\ &&
+\frac{i\,Z^2\,\alpha^3}{3\,\pi}
\,\biggl\langle\phi\biggl|
\left(\frac{\bfr_1}{r_1^3}+\frac{\bfr_2}{r_2^3}\right)\times 
\frac{(\bsigma_1+\bsigma_2)}{2}
\ln\left[\frac{2(H_0-E_0)}{(Z\,\alpha)^2}\right]\cdot
\left(\frac{\bfr_1}{r_1^3}+\frac{\bfr_2}{r_2^3}\right)
\biggr|\phi\biggr\rangle\,, 
\label{bethe}
\end{eqnarray}
\end{widetext}
where the term with the second order matrix element
has been dropped out, as all such terms are included in $E_S$. 
The term with $\ln(2\,\lambda)$ in Eq.~(\ref{bethe})
cancels out with the corresponding contribution in $H_Q$ in Eq. (\ref{hq}).

\section{Numerical evaluation and the results}
\label{sec4}

\subsection{Nonrelativistic wave function}

In order to obtain the nonrelativistic wave function, we use the technique
by Korobov \cite{korobov:99,korobov:00}, in which the spatial part of the triplet $P$
states is represented as
\begin{eqnarray} \label{wf}
\vec\phi(\bfr_1,\bfr_2) &=& \sum_{i=1}^N c_i 
\nonumber \\ && \times
\bigl[ \bfr_1
  \exp(-\alpha_i\,r_1-\beta_i\,r_2-\gamma_i\,r) 
- (1 \leftrightarrow 2) \bigr]\,,
\nonumber\\
\end{eqnarray}
where $r = |\bfr_1-\bfr_2|$. Real nonlinear parameters $\alpha_i$, $\beta_i$,
and $\gamma_i$ are chosen quasirandomly from the intervals
\begin{eqnarray} \label{wf1}
\alpha_i &\in& [A_1,A_2]\,, \nonumber \\
\beta_i &\in& [B_1,B_2]\,, \nonumber \\
\gamma_i &\in& [C_1,C_2]\,, 
\end{eqnarray}
with the parameters $A_{1,2}$, $B_{1,2}$, and $C_{1,2}$ being subjects of
a variational optimization. In order to enforce the proper
behavior of the wave function (\ref{wf}) in the limits $r_1\to\infty$,
$r_2\to\infty$, and $r\to\infty$, 
the nonlinear parameters are subjected to the condition
\begin{eqnarray}
\left\{ \alpha_i+\beta_i, \alpha_i+\gamma_i, \beta_i+\gamma_i\right\} >
  \sqrt{2 E_{\rm io}}\,, 
\end{eqnarray}
where $E_{\rm io}$ is the ionization energy of the atom. In order
to reproduce the behavior of the exact wave function for small
values of $r_{1,2}$ and $r$, the variational 
parameters $A_{1,2}$, $B_{1,2}$, and $C_{1,2}$ are
allowed to take negative values. 
To make the basis set more flexible, multiple  sets of the variational 
parameters $A_{1,2}$, $B_{1,2}$, and $C_{1,2}$ are introduced. 
Namely, the double basis set was used in this work for the determination of the
nonrelativistic wave function, and the triple basis set was used  
in calculations of corrections to the Bethe logarithm and 
second-order corrections. 

The calculation of matrix elements of the nonrelativistic Hamiltonian
is performed with the use of the simple formula for the master integral:
\begin{eqnarray}
\frac{1}{16\,\pi^2}\,\int d^3 r_1\,\int d^3r_2\,
\frac{e^{-\alpha r_1-\beta r_2-\gamma r}}{r_1\,r_2\,r} 
\nonumber \\&&\hspace*{-20ex}=
\frac{1}{(\alpha+\beta)(\beta+\gamma)(\gamma+\alpha)}.
\end{eqnarray}
Integrals with any additional powers of $r_i$ in the numerator can be 
obtained by differentiating with respect to the corresponding parameter 
$\alpha$, $\beta$, or $\gamma$. Matrix elements of relativistic corrections
involve additional inverse powers of $r_1$, $r_2$, and  $r$. They can be obtained 
by integrating with respect to the corresponding parameter.
In fact, all matrix elements required for the evaluation 
of the relativistic, QED, and the finite nuclear mass
corrections  can be expressed in terms of the rational, logarithmic, and
dilogarithmic functions of $\alpha,\beta$, and $\gamma$.

The procedure of generating the nonrelativistic wave function
looks now as follows. For the initial set of parameters $A_{1,2}$,
$B_{1,2}$, and $C_{1,2}$, the nonlinear parameters $\alpha_i$, $\beta_i$, and
$\gamma_i$  with $i=1,\ldots ,N$ are distributed quasirandomly. Then, 
the $N\times N$ matrix of the nonrelativistic Hamiltonian $H_0$ is
evaluated. The linear coefficients $c_i$ and the reference-state eigenvalue
$E_0$ are determined by using the inverse iteration method, with the LDU
decomposition employed for the inversion of the Hamiltonian matrix. Then the
procedure is repeated for a different set of parameters $A_{1,2}$,
$B_{1,2}$, and $C_{1,2}$, looking for the minimum value of the energy
$E_0$. The minimization problem is rather noisy, as the functional contains
many local minima. So, the simplest simplex-like algorithms of minimization
are probably the most appropriate for this task. Our calculations were
performed in the quadruple, sixtuple, and octuple arithmetics, which were
implemented in Fortran~95 by libraries written by V.~Korobov \cite{korobov:priv}. 

Our result for the nonrelativistic energy of helium for the triplet $P$ state
and the infinite nuclear mass is (in atomic units)
\begin{eqnarray} \label{energy}
E(2^3P) = -2.133\,164\,190\,779\,283\,205\,146\,96\,^{+0}_{-10}\,.
\end{eqnarray}
This value is the upper variational bound for the energy obtained with
$N=6600$ basis functions, and the uncertainty is the extrapolated lower bound.
The value in Eq. (\ref{energy}) is by about $4$ decimals more precise than
the previously best result of Ref. \cite{drake:02:cjp}. 

\subsection{Angular momentum algebra}
\label{sec:angular}

In the approach employed in the present investigation as well as in the
previous studies by K.P. and coauthors \cite{pachucki:00:jpb,pachucki:02:jpb:a},
all the angular momentum algebra is performed in Cartesian coordinates.
Tensor product of the $^3P$ wave functions is represented  \cite{forbidden} in terms
of the spatial wave functions and the spin operator
$\vec s = (\bsigma_1+\bsigma_2)/2$  as 
\begin{eqnarray}
|^3P_0\rangle\langle ^3P_0| &=& |i \rangle\langle j|\,
\biggl(\delta^{ij}\,\frac{s^2}{2}-s^j\,s^i\biggr)\,, \\
\frac{1}{3}\,\sum_m|^3P_1,m\rangle\langle ^3P_1,m| &=&
|i \rangle\langle j|\,\frac{1}{2}\,s^i\,s^j\,, \\
\frac{1}{5}\,\sum_m|^3P_2,m\rangle\langle ^3P_2,m| &=&
|i \rangle\langle j|\,
\frac{1}{10}\\ &&\times
\biggl(2\,s^2\,\delta^{ij}-3 s^i\,s^j+2\,s^j\,s^i\biggr)\,,\nonumber\\
\end{eqnarray}
where $|j\rangle$ denotes the state with the Cartesian index $j$ and
the normalization of the spatial wave functions is
\begin{equation}
\langle i|j\rangle=\delta^{ij}/3.
\end{equation}
For the calculation of the second order matrix elements
one needs formulae for the spin product
\begin{eqnarray}
s^i\, s^j\, s^k &=& \delta^{jk}\,s^i+
 \frac{i}{2}\,\epsilon^{jkl}\,s^i\,s^l+
 \frac{i}{2}\,\epsilon^{ikl}\,s^j\,s^l+
 \frac{i}{2}\,\epsilon^{ijl}\,s^k\,s^l\,,\nonumber \\
\end{eqnarray}
and for spin traces
\begin{eqnarray}
{\rm Tr}\,s^i &=& 0\,,\\
{\rm Tr}\,s^i\,s^j &=& 2\,\delta^{ij}\,,\\
{\rm Tr}\,s^i\,s^j\,s^k &=& i\,\epsilon^{ijk}\,,\\
{\rm Tr}\,s^i\,s^j\,s^k\,s^l &=&
\delta^{ij}\,\delta^{kl} + \delta^{jk}\,\delta^{il}\,.
\end{eqnarray}
Using these formulae, all matrix elements can be reduced to a form
involving the spatial wave functions only. For example, matrix elements of the
operators in the Breit-Pauli Hamiltonian can be expressed as
\begin{eqnarray}
\langle\vec Q\cdot\vec s\,\rangle_J &=& i\,\epsilon_{jkl}\,
\langle j|Q^k|l\rangle\,u_J\\
\langle\vec s\cdot\hat Q\cdot\vec s\rangle_J &=&
\langle j|Q^{jl}|l\rangle\,v_J
\end{eqnarray}
where $\hat Q$ is an arbitrary symmetric and traceless tensor ($Q^{jl}= Q^{lj}$ 
and $Q^{kk}=0$) and
\begin{eqnarray}
u_J &=& (1,1/2,-1/2),\\
v_J &=& (-1,1/2,-1/10)\,,
\end{eqnarray}
for $J=0,1,2$ respectively.

\subsection{Leading-order fine structure}

The dominant contribution to the fine structure comes from the spin-dependent
part of the Breit-Pauli Hamiltonian $H_{\rm fs}$ given by Eq.~(\ref{fs}). With 
including the nuclear recoil effect, the leading-order contribution to the
fine structure in helium is
\begin{eqnarray}
E_{\rm fs}(J) &=& \langle H_{\rm fs}\rangle_J\nonumber \\
       &=&\left({m_r \over m}\right)^3 \frac{\alpha^4}{4} \biggl[ 
           -E_1\,(1+a_e)^2\,v_J 
           +E_2\,(1+2\,a_e)\,u_J\nonumber\\ &&
           +E_3\,\biggl(1+\frac{4}{3}\,a_e\biggr)\,u_J
           +\frac{m}{M}\,E_4\,(1+a_e)\,u_J \biggr]\,.
\end{eqnarray}
The corresponding results for the large and the small fine structure interval are
\begin{eqnarray} \label{lo1}
\nu_{01} & = &  \left({m_r \over m}\right)^3 
\alpha^2 R_{\infty}c\biggl[ \frac{3 E_1}{4}(1+a_e)^2+ \frac{E_2}{4}(1+2a_e) \nonumber \\ 
& + & \frac{E_3}{4}(1+ {4 \over 3}a_e)+{m \over M}\frac{E_4}{4}(1+a_e)\biggr]\,,  \\
\nu_{12} & = & \left({m_r \over m}\right)^3 \alpha^2 R_{\infty}c\biggl[ -\frac{3 E_1}{10}(1+a_e)^2+
\frac{E_2}{2}(1+2a_e) \nonumber \\
& + & \frac{E_3}{2}(1+ {4 \over 3}a_e)+{m \over
  M}\frac{E_4}{2}(1+a_e)\biggr]\,,
 \label{lo2}
\end{eqnarray}
where the constants $E_i$ are given by (in atomic units)
\begin{eqnarray}
E_1 &   =&  2\,\biggl\langle j\biggl|
            3\,\frac{r^j r^i}{r^5} - \frac{\delta^{ji}}{r^3}
            \biggr|i\biggr\rangle\,, \\
E_2 &   =&  2\,Z\,\epsilon_{jki}\,\biggl\langle j\biggl|
             \biggl(\frac{\bfr_1}{r_1^3}\times\vec\nabla_1\biggr)^k
             \biggr|i\biggr\rangle, \\
E_3 &   =&  -3\,\epsilon_{jki}\,\biggl\langle j\biggl|
            \biggl(\frac{\bfr}{r^3}\times\bigl(\vec\nabla_1-\vec\nabla_2\bigr)\biggr)^k
            \biggr|i\biggr\rangle, \\
E_4 &   =&  4\,Z\,\epsilon_{jki}\,\biggl\langle j\biggl|
            \biggl(\frac{\bfr_1}{r_1^3}\times\bigl(\vec\nabla_1+\vec\nabla_2\bigr)\biggr)^k
            \biggr|i\biggr\rangle\,.
\end{eqnarray}
Our numerical results for the constants $E_i$ are
\begin{eqnarray}
E_1 &   =& ~~~0.180~220~618~632~744\,(10) \,, \\
E_2 &   =&   -0.277~401~358~712~829\,(10) \,, \\
E_3 &   =& ~~~0.411~999~963~626~094\,(25) \,, \\
E_4 &   =& ~~~0.241~945~125~695~21\,(6) \,.
\end{eqnarray}
These results are obtained with including the mass polarization term
into the zeroth-order Hamiltonian and thus contain effects of the second and
higher orders in $m/M$. 

\subsection{$\bm{m}\,\bm{\alpha}^{\bf 6}$ contribution}

The contribution of order $m\,\alpha^6$ to the fine structure of helium
is represented by Eq.~(\ref{fs6}). The most difficult part of its numerical
evaluation is associated with the second-order contributions.
First calculations of the second-order corrections to the helium fine
structure were performed by Hambro \cite{hambro:72} and Lewis and
Serafino \cite{lewis:78}. Two decades later, Yan and Drake \cite{yan:95:prl}
did these calculations to a much higher accuracy and demonstrated that the
first results were much less accurate than it was claimed. The
nuclear recoil and the amm effects were included into the second-order
corrections in Ref.~\cite{drake:02:cjp}. An independent evaluation of the
second-order corrections (including the amm part but not the recoil effect) was
performed by K.P. and Sapirstein \cite{pachucki:02:jpb:a}. In the present
work, we re-calculate all $m\,\alpha^6$ corrections, with the intention
to independently check the numerical convergence of the previous results
and, more importantly, to check the nuclear recoil effect on the second-order 
corrections, which was previously calculated only by Drake. 

The second-order corrections in Eq.~(\ref{fs6}) are of two kinds: the
symmetric and the non-symmetric one. The symmetric contributions are the most
numerous ones since they involve the $^3P$, $^1P$, $^3D$, $^1D$, and $^3F$ 
intermediate states. The derivation of the calculational formulas for them is
relatively straightforward along the lines presented in Sec.~\ref{sec:angular}.

The numerical evaluation of the symmetric second-order contributions was
performed by employing the variational optimization of the nonlinear parameters
of the basis set for the Green function. Convergence of numerical results
is rather slow for the $^3P$ intermediate states because of the
singular character of the Breit interaction. The convergence can be improved
by introducing singular functions into the basis set, as in
Ref.~\cite{yan:95:prl}. We, however, prefer to exploit the flexibility of the
basis set (\ref{wf}) and emulate the missing basis functions by using
very large exponents. In order to effectively span large regions of
nonlinear parameters, we used non-uniform distributions 
of the kind \cite{pachucki:02:jpb}
\begin{equation}
\alpha_i = A_1+ (t_i^{-a}-1)A_2\,,
\end{equation}
with $a = 2$ and 3, where the variable $t_i$ has a uniform quasirandom distribution over the
interval $(0,1)$.

The non-symmetric second-order contributions involve only the $^3P$
intermediate states. Care should be taken in the numerical calculation of this
part due to the presence of singular operators, namely the Dirac $\delta$ and
the $\bfp^{\,4}$ operators. While a straightforward numerical evaluation is
possible, a much better convergence is obtained by transforming the singular
operators to a more regular form. 

The treatment of the second-order correction involving the $\delta$
operator is based on the global representation of the $\delta$ function
introduced by Drachman \cite{drachman:81},
\begin{align} 
\langle 0|4\pi&\,\delta^3(r_1)|n\rangle 
\nonumber \\ &
= \lbr 0\left| \frac{2}{r_1}(f_n+f_0) 
 +\sum_{a=1,2} \bnabla_a\cdot  \frac{2}{r_1}\,\bnabla_a \right| n\rbr\,,
\nonumber \\
\end{align}
where 
\begin{equation}
f_k = E_k -V + \frac{m_r}{M}\,\bnabla_1\cdot\bnabla_2\,,
\end{equation}
and $V = -Z/r_1-Z/r_2+1/r$.
Noting that $f_n = E_n-E_0+f_0$, one can cancel the $E_n-E_0$ factor 
with the denominator of the reduced Green function in the second-order matrix 
element. Using the completeness
of the eigenfunctions, we obtain the regularized
expression for the second-order correction,
\begin{align}
&\left< 4\pi \left[ \delta^3(r_1)+\delta^3(r_2)\right]\,
  \frac1{(E_0-H_0)^{\prime}}\, H_{\rm fs}\right>
\nonumber \\ & \ \ \ \ 
= \left< \left( \frac1{r_1}+\frac1{r_2}\right)
  \, 4f_0\, \frac1{(E_0-H_0)^{\prime}}\, H_{\rm fs}\right>
 \nonumber \\ & \ \ \ \ 
+\sum_{a=1,2}  \lbr \bnabla_a\cdot
           \left( \frac2{r_1}+\frac2{r_2}\right)
           \bnabla_a\, \frac1{(E_0-H_0)^{\prime}}\, H_{\rm fs}\right>
 \nonumber \\ & \ \ \ \ 
 -\left<  \left(\frac2{r_1}+\frac2{r_2} \right)\,  H_{\rm fs}\right>
 +\left<  \frac2{r_1}+\frac2{r_2}\right> \lbr  H_{\rm fs}\rbr\,.
\end{align}

A regularized expression for the second-order correction with the
operator $-1/8(\bfp_1^{\,4}+\bfp_2^{\,4})$ can be derived by 
using the identity
\begin{equation}
-\frac18\,(\bfp_1^{\,4}+\bfp_2^{\,4}) = -\frac18\,(\bfp_1^{\,2}+\bfp_2^{\,2})^2
+ \frac14\, \bfp_1^{\,2} \bfp_{2}^{\,2}
\end{equation}
and employing the Schr\"odinger equation to transform the first term, as described in
Ref.~\cite{drake:02:cjp}. 

For the numerical evaluation of the non-symmetric contributions we used the
set of nonlinear parameters obtained by merging two subsets, one obtained by
the variational optimization of the symmetric second-order Breit correction and
another, by an optimization of the symmetric correction for the model
perturbation $\delta V = 1/r_1^2+1/r_2^2$. The model potential $\delta V$
corresponds to the most singular part of the regularized $\delta$ operator.

%
%
\begin{table*}
\caption{Second-order contributions to order $m\,\alpha^6$ for the large
($\nu_{01}$) and small ($\nu_{12}$) fine-structure intervals in helium, including the
recoil and the amm corrections. Units are kHz.
\label{tab:second} }
\begin{ruledtabular}
\begin{tabular}{c|...|...}
     State        & \multicolumn{3}{c}{$\nu_{01}$}
                               & \multicolumn{3}{c}{$\nu_{12}$} \\
      & &  \multicolumn{1}{c}{recoil} & \multicolumn{1}{c}{amm}
      &  &  \multicolumn{1}{c}{recoil} & \multicolumn{1}{c}{amm} \\
\hline\\[-9pt]

$^3P$ &  -4894x.29(2)  &  -1x.09 & -14x.40  &   -1569x.62  &    0x.21 &  -3x.20\\
$^1P$ &   6595x.64     &  -9x.29 &  23x.28  &   -6595x.64  &    9x.29 & -23x.28\\
$^3D$ &     26x.33     &  -0x.01 &   0x.07  &      50x.50  &   -0x.03 &   0x.16\\
$^1D$ &                &         &          &      22x.20  &    0x.05 &   0x.05\\
$^3F$ &                &         &          &      52x.24  &   -0x.01 &   0x.24\\
Sum   &   1727x.68(2)  & -10x.39 &   8x.96  &   -8040x.32  &    9x.51 & -26x.02\\
Drake~{\cite{drake:02:cjp}}&
          1727x.58(4)  & -10x.81(4)& 8x.95  &   -8040x.38(5)&  10x.19(11)& -26x.02\\
\end{tabular}
\end{ruledtabular}
\end{table*}

The numerical results for the $m\,\alpha^6$ second-order corrections are
presented in Table~\ref{tab:second}. Our values for these corrections
in the non-recoil limit (both with and without the amm part) agree
well with the results by Drake \cite{drake:02:cjp}. In the recoil part of the
second-order corrections, we observe some deviation from Drake's results. 
To localize the source of the discrepancy, we separate the recoil contributions
into 3 parts, which are unduced by 
the mass scaling, the mass polarization, and the recoil
operators. The corresponding contributions to the $\nu_{01}$ interval from the
$^3P$ intermediate states are $3.26 -3.66 -0.69 = -1.09$ kHz, to be compared
with Drake's values of $3.25-3.66-0.06 = -0.47$ kHz. Analogous contributions
to the $\nu_{12}$ interval are $1.17 -1.87+ 0.91 = 0.21$ kHz, to be compared
with Drake's values of $1.17-2.15+0.91 = -0.07$ kHz. The contributions to 
the $\nu_{01}$ interval from the
$^1P$ intermediate states are $-4.54 -7.68+ 2.93= -9.29$ kHz, to be compared
with Drake's values of $-4.52 -7.66+ 1.87= -10.31$ kHz. 
The total difference between our results and those by Drake is rather small
numerically and does not influence significantly the
comparison of theory with the experimental data.

\subsection{Relativistic correction to the Bethe logarithm}

The relativistic correction to the Bethe logarithm is given by Eq.~(\ref{bethe}).
The two $\lambda$-independent
parts in the right-hand side of this equation will be referred to as the $E_{L1}$ and
$E_{L2}$ corrections, respectively. We will start our discussion 
with the simpler part $E_{L2}$. 
For the numerical evaluation, it is convenient to transform 
this correction to the equivalent form,
\begin{eqnarray} \label{L21}
E_{L2} = -\frac{\alpha}{3\,\pi}\int_0^\infty dk\,\biggl[k^2\,L_2(k)-A_2\biggr]\,,
\end{eqnarray}
where
\begin{eqnarray} \label{L22}
L_2(k) &=& -i\lbr (\bfp_1+\bfp_2)\times
\frac{1}{H_0+k-E_0}\,(\bfp_1+\bfp_2)\cdot \vec{s}\rbr\,, \nonumber \\
\end{eqnarray}
and $A_2$ is the leading term of the large-$k$ asymptotic expansion of
$k^2L_2(k)$, which has the form
\begin{equation} \label{L23}
k^2\,L_2(k) =
A_2 +\frac{B_2}{k^{3/2}} + \frac{C\,\ln k}{k^2}+\frac{D}{k^2}+
\frac{E}{k^{5/2}}+\ldots\,.
\end{equation}
The two leading asymptotic constants are evaluated to be
\begin{eqnarray} \label{L24}
A_2 &=& Z\,\lbr \biggl(\frac{\bfr_1}{r_1^3}+
\frac{\bfr_2}{r_2^3}\biggr)\times
(\bfp_1+\bfp_2)\cdot \vec{s} \rbr\,, \\
B_2 &=& -i\,\frac{4\,\pi\,Z^2}{3\,\sqrt{2}}\,
\lbr \bfp_1\times \delta^3(r_1)\,\bfp_1\cdot\vec{s}+
        \bfp_2\times \delta^3(r_2)\,\bfp_2\cdot \vec{s}\rbr\,. \nonumber \\
\end{eqnarray}
Here we correct the overall sign in Eqs.~(\ref{L22}) and (\ref{L24})
as compared to Ref.~\cite{pachucki:00:jpb}. It is
noteworthy that the $k^{-2}$ asymptotic
behavior of $L_2(k)$ arises through an internal cancellation of the three angular-momentum
contributions ($^{3}S^e$, $^{3}P^e$, and $^{3}D^e$, where $e$ stands
for the even parity), since each of them
separately falls off as $k^{-1}$ only.

In order to accurately perform the integration in Eq. (\ref{L21}), we transform 
this expression to the following form
\begin{eqnarray} \label{L25}
E_{L2} &&= -\frac{\alpha}{3\,\pi}
 \Biggl\{ \int_0^{\kappa} dk\,k^2\,L_2(k) 
\nonumber \\ && 
+ \int_{\kappa}^{\infty}dk\,
\left[k^2\,L_2(k)-A_2-\frac{B_2}{k^{3/2}}\right] 
-A_2\kappa+ \frac{2B_2}{\sqrt{\kappa}}\Biggr\}\,,  
 \nonumber \\
\end{eqnarray}
where $\kappa$ is a free parameter. 

In the numerical evaluation of $E_{L2}$, we exploit the fact that
the integrand $L_2(k)$ obeys the variational principle, similarly to 
that for the Bethe logarithm \cite{Schwartz:61}. In fact, each angular-momentum
contribution to $L_2(k)$ has the same form as for the Bethe logarithm, the
difference being only the prefactors coming from the angular-momentum
algebra. (It is important that the difference in the prefactors leads to the
disappearance of the $k^{-1}$ term in the large-$k$ asymptotics of $L_2(k)$.)
In order to perform the integration over $k$ in Eq.~(\ref{L25}), one needs to
know the function $L_2(k)$ for a wide region of $k$. As noted in
Ref.~\cite{komasa:01}, there is no need to perform the full variational
optimization of the basis for each value of $k$. The idea is that, having
got the optimized set of nonlinear parameters for the basis at $k=k_1$
and $k=k_2$, for all $k$ in between one can use the basis obtained by merging
together the two optimized sets. 
The asymptotic behavior of the integrand $L_2(k)$ for large $k$, together with its value at
$k=0$, $L_2(0) = -\langle(\vec r_1+\vec r_2)\times(\vec p_1+\vec p_2)\cdot\vec
s\rangle$, served as useful tests of the numerical procedure. 

The general evaluation scheme is as follows. First, we perform a careful
optimization of nonlinear basis-set parameters for several distinct
scales of $k$: $k_i = 10^i$, with $i = 1,\ldots,i_{\rm max}$ and $i_{\rm max}=4$. 
The optimization is carried out with incrementing the size of the
basis, until the prescribed accuracy is achieved. The size of the optimized
basis employed in actual
calculations varied from $N=600$ for $k_1 = 10$ to $N=1600$ for the $^{3}D^e$
wave and $k_4 = 10^4$, yielding the numerical accuracy of about 10 digits
for $L_2(k)$. For each particular value of $k\leq 10^{i_{\rm max}}$, the
calculational basis is obtained by merging the optimized bases for the two 
closest $k_i$ points, thus essentially doubling the number of the basis
functions. According to our experience, such merging usually yields an
additional digit of accuracy. 

The integral over $k\in[0,\kappa]$ in Eq.~(\ref{L25}) was calculated
analytically, after performing the full diagonalization of the Hamiltonian
matrix and using the spectral representation of the propagator. This allowed us
to avoid problems associated with the pole on the real axis coming from the $2^3S$ state. 
The parameter $\kappa$ was set to $\kappa=10$. The
integral over $k\in[\kappa,\infty)$ was separated into two parts, $k< 10^{i_{\rm max}}$
and $k> 10^{i_{\rm max}}$. The first part was evaluated by using the
Gauss-Legendre quadratures, after the change of variables $t = 1/k^2$. The
second part was evaluated by fitting the integrand to the
form
\begin{eqnarray} \label{fit}
k^2\,L_2(k)-A_2-\frac{B_2}{k^{3/2}} = \frac{\ln k}{k^2}\,W_1\left(\frac1k\right) +
    \frac{1}{k^2} \, W_2\left(\frac1{\sqrt{k}}\right),\nonumber \\
\end{eqnarray}
where $W(x)$ denotes a polynomial of $x$. For fitting, we used the function
$L_2(k)$ stored on the interval $k=1,\ldots,100$. The total number of fitting
parameters in the above expression was about $9-11$. The optimal form of the
fitting function was selected by demanding it to reproduce the known
asymptotic constants $A_2$ and $B_2$ for the function $L_2(k)$. The error due to the
fitting procedure was estimated by comparing the integration results for 
the fitted function and for the numerical integrand outside the fitting
region, i.e., for $k\in [10^2,10^4]$. 

Our results for the asymptotic constants $A_2$ and $B_2$ are
\begin{eqnarray} 
A_2 &=&  0.120~944~339~354~433\,(8)\,u_J\,, \\
B_2 &=& -0.982~581~108\,(2)\,u_J\,.
\end{eqnarray}
The final result for the $E_{L2}$ correction to the helium fine structure is
(in units $m\,\alpha^7$)
\begin{eqnarray} \label{L2num}
E_{L2} = 0.067~682~1(5)\,u_J\,,
\end{eqnarray}
This is in reasonable agreement with the
value obtained previously in Refs.~\cite{pachucki:00:jpb,pachucki:02:jpb:a},
which is $-0.06775(5)\,u_J$, except for the overall sign, which we correct here.

We now turn to the evaluation of the $E_{L1}$ correction.
In terms of the integral over the photon momentum, it is written as
\begin{eqnarray} \label{L11}
E_{L1} = -\frac{2\alpha}{3\,\pi} \lim_{K\to\infty} 
 \Biggl[
  \int_0^K dk\,k\,L_1(k) - A_1\,K -B_1\,\ln K \Biggr]\,,  \nonumber \\
\end{eqnarray}
where 
\begin{align} \label{L12}
L_1(k)   &\ = 2\,\biggl\langle  H^{(4)}_{\rm fs}\,\frac{1}{(E_0-H_0)'}\,
(p_1^i+p_2^i)\,\frac{1}{H_0+k-E_0}
\nonumber \\ & \times
(p_1^i+p_2^i)\biggr\rangle + 
\biggl\langle(p_1^i+p_2^i)\,\frac{1}{H_0+k-E_0} 
\nonumber \\ &\times
\biggl[\langle H^{(4)}_{\rm fs}\rangle -H^{(4)}_{\rm fs}\biggr]\,
\frac{1}{H_0+k-E_0}\,(p_1^i+p_2^i)\biggr\rangle\,,
\end{align}
and $A_1$ and $B_1$ are the leading terms of the large-$k$ asymptotic
expansion of the integrand,
\begin{equation} \label{L13}
k\,L_1(k) = 
A_1+\frac{B_1}{k}+\frac{C}{k^{3/2}}+\frac{D\,\ln k}{k^2}+\frac{E}{k^2}
+\ldots\,,
\end{equation}
with
\begin{eqnarray}
A_1 &=& 2\,\lbr H^{(4)}_{\rm fs}\,\frac{1}{(E_0-H_0)'}
(\bfp_1+\bfp_2)^2\rbr\,, \\
B_1 &=& -\lbr H^{(4)}_{\rm fs}\,\frac{1}{(E_0-H_0)'}
\,4\,\pi\,Z\,[\delta^3(r_1)+\delta^3(r_2)]\rbr\nonumber \\ &&
- \frac{i\,\pi\,Z}{2}\,\lbr 
\bfp_1\times\delta^3(r_1)\,\bfp_1\cdot\vec{s}+
\bfp_2\times\delta^3(r_2)\,\bfp_2\cdot\vec{s}
\rbr\,. \nonumber \\
\end{eqnarray} 
At $k=0$, the integrand $L_1$ can be evaluated analytically to yield $L_1(0) = 0$. 

For the numerical evaluation, Eq.~(\ref{L11}) is written in the form similar
to Eq.~(\ref{L25}),
\begin{eqnarray} \label{L15}
E_{L1} &&= -\frac{2\alpha}{3\,\pi}
 \Biggl\{ \int_0^{\kappa} dk\,k\,L_1(k) 
\nonumber \\ &&  \!\!\!\!\!\!\!
+ \int_{\kappa}^{\infty}dk\,
\left[k\,L_1(k)-A_1-\frac{B_1}{k}\right] 
-A_1\kappa- B_1\,\ln\kappa\Biggr\}\,.
 \nonumber \\
\end{eqnarray}
The main difference of the numerical evaluation of $E_{L1}$ from that of
$E_{L2}$ is that the integrand 
$L_1(k)$, contrary to $L_2(k)$, does not obey the variational
principle [i.e., there is no functional whose minimum yields the
exact value of $L_1(k)$]. Because of this, in evaluation of $E_{L1}$
we have to use the variational
optimization results for the nonlinear basis-set parameters obtained for
$E_{L2}$. This is a serious drawback since it is clear that the optimal
set of parameters for the integrand $L_2(k)$ is not exactly optimal for
$L_1(k)$, because of an additional singularity introduced by the perturbing
Hamiltonian $H^{(4)}_{\rm fs}$. After some numerical experimenting, we found that this
additional singularity can be well accounted for if the calculational basis
for each $k$ is not just doubled by merging two sets optimized for two scales
$k_1$ and $k_2$, but tripled, with the third part obtained from the second one
by (alternatively) scaling the parameters $\alpha_i$ and $\beta_i$ 
by a factor $g = 10$. This
trick was inspired by the method described in Ref.~\cite{korobov:04}. 

With this modification, our numerical evaluation of $E_{L1}$ was done similarly to
that for $E_{L2}$. Because of the tripling of the basis set, we used the
optimized parameters with somewhat smaller number of the basis functions but
increased the high-energy cutoff parameter up to $k_5 = 10^5$. Our results for
the asymptotic constants $A_1$ and $B_1$ are 
\begin{eqnarray}
A_1 &=& -0.028\,038\,047\,8\,(10)\,u_J + 0.054\,037\,866\,(4)\,v_J \,, \nonumber \\ \\
B_1 &=& -0.169\,127\,85\,(20)\,u_J + 0.146\,477\,680\,(2)\,v_J\,.
 \nonumber \\
\end{eqnarray}
The final result for the $E_{L1}$ correction to the helium fine splitting is
(in units $m\,\alpha^7$)
\begin{eqnarray}
E_{L1} = -0.107\,664\,(6)\,u_J + 0.118\,404\,4\,(4)\,v_J\,.
\end{eqnarray}
The spin-orbit part of the above result is by about $25\%$ larger than
the previously reported value of Ref.~\cite{pachucki:00:jpb} of
$-0.0817(20)u_J$, whereas the spin-spin part is by about $10\%$ larger
that the previous value of $0.0959(4)v_J$. The reason for this deviation lies
in the insufficient accuracy of the previous calculations.


\section{Summary and discussion}
\label{sec:summary}

%
%
\begin{table}
\caption{Summary of individual contributions to the helium fine structure. 
Units are kHz.
\label{tab:summary} }
\begin{ruledtabular}
\begin{tabular}{l..}
            & \multicolumn{1}{c}{$\nu_{01}$}
                               & \multicolumn{1}{c}{$\nu_{12}$} \\
\hline\\[-9pt]
  $E^{(4)}$           &   29~618~418x.54      &    2~297~717x.82   \\
  $E^{(6)}$           &       -1~556x.97(2)   &       -6~544x.93   \\
  $E^{(7)}$[log]      &           82x.59      &          -10x.09   \\
  Subtotal            &   29~616~944x.16(2)  &     2~291~162x.80 \\
 Drake~\cite{drake:02:cjp}
                      &   29~616~943x.40(6)  &     2~291~163x.40(13) \\ 
  $E_Q$[nolog]        &          21x.73      &            7x.42   \\
  $E_H$               &          -4x.21      &            4x.05   \\
  $E_S$[nolog]        &          11x.42      &           -1x.21   \\
  $E_{L1}$            &          -31x.51      &           -4x.99   \\
  $E_{L2}$            &            4x.61      &            9x.22   \\
  $E^{(7)}$[nolog]    &            2x.04      &           14x.48   \\
  Higher orders       &        \pm1x.6       &         \pm1x.6    \\ 
  Total theory        &   29~616~946x.20\pm1.6   &    2~291~177x.28\pm1.6   \\
  Experiment \cite{borbely:09} &                &    2~291~177x.53(35) \\
  Experiment \cite{zelevinsky:05}   
                      &   29~616~951x.66(70)    &    2~291~175x.59(51) \\   
  Experiment \cite{george:01}   
                      &   29~616~950x.9(9) \\
  Experiment \cite{giusfredi:05}
                      &   29~616~952x.7(1.0)    &    2~291~168x.(11.) \\   
\end{tabular}
\end{ruledtabular}
\end{table}

The summary of all contributions available for the fine structure of helium is
given in Table~\ref{tab:summary}. Numerical results are presented for the
large $\nu_{01}$ and the small $\nu_{12}$ intervals,
defined by
\begin{eqnarray}
\nu_{01} = \bigl[ E(2^3P_0)-E(2^3P_1)\bigr]/h\,, \\
\nu_{12} = \bigl[ E(2^3P_1)-E(2^3P_2)\bigr]/h\,. 
\end{eqnarray}
The parameters used in our calculations are: $\alpha^{-1} = 137.035~999~679(94)$, $cR_{\infty}
= 3~289~841~960~361(22)$~kHz, and $m/M = 1.370~933~555~70 \times 10^{-4}$. 
In the table, the correction $E^{(4)}$ is given by Eqs.~(\ref{lo1}) and
(\ref{lo2}) and the correction $E^{(6)}$, by Eq.~(\ref{fs6}). $E^{(7)}$[log] 
denotes the sum of the logarithmic parts of $E_S$ and $E_Q$. 
The corrections $E_Q$, $E_H$, and
$E_S$ are given by Eqs.~(\ref{HQ}), (\ref{EH}), and (\ref{ES}), respectively. 
The complete listing of numerical results for individual terms contributing to 
$E_Q$ and $E_H$ can be found in Ref.~\cite{pachucki:06:prl:he} and is not
repeated here. The
relativistic corrections to the Bethe logarithm $E_{L1}$ and $E_{L2}$ are
given by Eqs.~(\ref{L11}) and (\ref{L21}), respectively.

The result for $E^{(4)}$ in Table~\ref{tab:summary} is consistent
with that of Ref.~\cite{pachucki:06:prl:he} after accounting for the newer value of
the fine structure constant. 
The result for $E^{(6)}$ differs slightly from the corresponding value
in Ref.~\cite{pachucki:06:prl:he}, mainly because of the change in the recoil
second-order Breit correction, which was previously calculated only by Drake
\cite{drake:02:cjp}.  As can be seen from the table, different 
theoretical predictions that
include contributions up to order $m\,\alpha^7 \log \alpha$ and $m^2/M\,\alpha^6$
(entry ``Subtotal'') agree at a sub-kHz level with each other. 

The nonlogarithmic correction to order $m\,\alpha^7$ has not been
checked independently. The compilation of results presented  for this
correction in Ref.~\cite{drake:02:cjp} is in part based on the
derivation by Zhang \cite{zhang:96:a,zhang:96:b,zhang:97}, which was shown to
be not entirely consistent \cite{pachucki:06:prl:he}, and in part includes
calculational results by K.P. and Sapirstein \cite{pachucki:00:jpb}. 

Table~\ref{tab:summary} shows that the calculational error of our
results is almost negligible as compared to the experimental uncertainty. 
There is, however, a much larger
theoretical error induced by the higher-order corrections.
It was believed previously \cite{zhang:96:prl,pachucki:00:jpb} that the
higher-order $m\,\alpha^8$ effects contribute well under the 1~kHz level. 
Particularly, the analysis presented in Ref.~\cite{pachucki:00:jpb} identified
several $m\,\alpha^8$ corrections that are enhanced by $\ln (Z\alpha)$ but nevertheless
contribute only about 0.1~kHz. In the present investigation, we found several
nonlogarithmic corrections that might contribute at the 1~kHz level. 

The first contribution comes from the mixing between the $^3P_1$ and $^1P_1$ levels. 
The nonrelativistic $2^1P_1-2^3P_1$ energy difference of $61.3\times 10^6$ MHz
acquires the relativistic correction of $-17.1\times 10^3$ MHz.
If we consider the $m\,\alpha^6$ second-order Breit correction to the energy
of the $2^3P_1$ state with the $2^1P_1$ intermediate states, the modification of
the $2^1P_1-2^3P_1$ energy difference by relativistic effects alters the
value of the correction by about 1 kHz. We thus estimate the theoretical uncertainty of the
$\nu_{01}$ and $\nu_{12}$  fine structure intervals due to the mixing between
the $^3P_1$ and $^1P_1$ levels as $\pm 1.0$ kHz. It should be mentioned, however, that all
corrections due to the mixing cancel identically in the sum of the large and 
small intervals, $\nu_{02} = \nu_{01}+\nu_{12}$. It is, therefore, likely that
the theoretical value for the interval $\nu_{02}$ is more
accurate than that for the interval $\nu_{01}$ and $\nu_{12}$ separately. 

The largest identified $m\,\alpha^8$ contribution to the interval $\nu_{02}$ 
comes from the one-photon
exchange diagram, which was evaluated to all orders in $Z\alpha$ but to the
leading order in $1/Z$ in Ref.~\cite{mohr:85:pra}. The result obtained in
that work for $\nu_{02}$ is $0.1033\,m\,\alpha^8\,Z^7$. 
Because of the $Z^7$ enhancement, the numerical contribution for helium is
quite large, $13.1$~kHz. This is, however, only the
leading term of the $1/Z$ expansion;
the complete contribution for helium is going to be much smaller
because of the screening. In order to estimate the screening effect, we
compare the complete contribution to order $m\,\alpha^6$ for $\nu_{02}$ in
helium, which is $-8.11$~MHz, with the corresponding one-photon exchange term,
which gives $86.4$~MHz. The resulting estimate is $\pm 1.2$~kHz. 

The total theoretical error due to 
the higher-order effects specified in Table~\ref{tab:summary} for 
the $\nu_{01}$ and $\nu_{12}$ intervals is obtained by
adding quadratically the two error estimates discussed above. We observe that for the small
interval, the theoretical value agrees well with the experimental results, whereas
for the large interval, a disagreement of about 3 standard deviations is
present. It should be noted that the present theoretical uncertainties are
much larger than those specified in previous investigations, the reason being that
in most previous cases, the uncertainties represented the calculational errors only.

Commenting on the situation when theory agrees with experiment for one
fine-structure interval and disagrees for another, we have to state that 
we do not have any satisfactory explanation for it.
All effects contributing to one interval contribute also to the other, both 
contributions being comparable in magnitude. We thus see no reason why a
theoretical prediction for one interval should be significantly more accurate
than for the other. Presuming that the
experimental value for the $\nu_{01}$ interval is correct, we have to conclude
that the excellent agreement of our theoretical value for the $\nu_{12}$ interval
with the latest measurement by Borbely {\em et al.}~\cite{borbely:09} is probably
accidental. 

Finally, we present separately the theoretical result for the sum of the large
and the small fine structure intervals in helium, $\nu_{02}({\rm theo}) =
31\,908\,123.5\,(1.2)$~kHz. For the reason discussed above, its uncertainty is
smaller than for the $\nu_{01}$ and $\nu_{12}$ intervals. The
theoretical value disagrees with the experimental result of
$\nu_{02}({\rm exp}) = 31\,908\,126.78\,(94)$~kHz \cite{zelevinsky:05} by
about 2 standard deviations. 

To conclude, we performed an evaluation of the helium fine structure that is complete
to orders $m\,\alpha^7$ and $m^2/M\,\alpha^6$. Our results for the
$m\,\alpha^4$, $m\,\alpha^5$, and $m\,\alpha^6$ contributions 
agree with those reported in previous investigations at a sub-kHz level. The
present evaluation of the relativistic corrections to the Bethe logarithm 
significantly improves upon the original calculation. The corresponding results reduce the
previously reported discrepancy between the theoretical predictions and the experimental
results. However, the remaining difference for the $\nu_{01}$ interval
is larger than the estimated
contribution of the higher-order effects. This discrepancy needs to be
resolved in order to make possible the determination of the fine structure
constant by means of the helium spectroscopy.

\section*{Acknowledgments}

This work was supported by NIST through Precision Measurement Grant PMG 60NANB7D6153.
V.A.Y. acknowledges additional support from the ``Dynasty'' foundation and from RFBR 
(grant No.~06-02-04007).


\appendix

\section{Foldy-Wouthuysen transformation in $\bm{d}$-dimensions}
\label{appendix}

The Foldy-Wouthuysen (FW) transformation \cite{itzykson:80} is the
nonrelativistic expansion of the Dirac Hamiltonian in an external
electromagnetic field. Following Ref. \cite{pachucki:06:hesinglet} 
we extend this transformation to 
the case where the dimension $d$ of space is arbitrary.
The Dirac Hamiltonian in an external electromagnetic field is
\begin{equation}
H = \vec \alpha \cdot \bpi +\beta\,m + e\,A^0\,,
\end{equation}
where $\bpi = \bfp-e\,\bfA$,
\begin{equation}
\alpha^i = \left(
\begin{array}{cc}
0&\sigma^i\\
\sigma^i&0
\end{array}\right),\;\;
\beta = 
\left(
\begin{array}{cc}
I&0\\
0&-I
\end{array}\right),
\end{equation}
and
\begin{equation}
\{\sigma^i,\sigma^j\} = 2\,\delta^{ij}\,I.
\end{equation}
The FW transformation $S$ \cite{itzykson:80} leads to a new Hamiltonian
\begin{equation}
H_{\rm FW} = e^{i\,S}\,(H-i\,\partial_t)\,e^{-i\,S}\,, 
\end{equation}
which decouples the upper and the lower component of the Dirac wave
function up to a specified order in the $1/m$ expansion. Here 
we calculate $H_{\rm FW}$ up to terms contributing
to the $m\,\alpha^6$ correction to the energy.  We use a convenient
form of the FW operator $S$, which can be written as 
\begin{eqnarray}
S &=&-\frac{i}{2\,m}\,\biggl\{ \beta\,\vec\alpha\cdot\bpi -
\frac{1}{3\,m^2}\,\beta\,(\vec\alpha\cdot\bpi)^3
\nonumber \\ && 
+\frac{1}{2\,m}\,[\vec\alpha\cdot\bpi\, ,\,e\,A^0-i\,\partial_t]
+\frac{\beta}{5\,m^4}\,(\vec\alpha\cdot\bpi)^5 
\nonumber \\ && 
-\frac{\beta\,e}{4\,m^2}\,\vec\alpha\cdot\dot{\bfE}+
     \frac{i\,e}{24\,m^3}\,[\vec\alpha\cdot\bpi,
            [\vec\alpha\cdot\bpi,\vec\alpha\cdot\bfE]]
\nonumber \\ && 
     -\frac{i\,e}{3\,m^3}\,\bigl\{(\vec\alpha\cdot\bpi)^2\,,\,
            \vec\alpha\cdot\bfE\bigr\}\biggr\}\,.
\end{eqnarray}
The FW Hamiltonian is expanded in a power series in $S$
\begin{equation}
H_{\rm FW} = \sum_{j=0}^6 {\cal H}^{(j)}+\ldots \label{b06}
\end{equation}
where 
\begin{eqnarray}
{\cal H}^{(0)} &=& H, \nonumber \\
{\cal H}^{(1)} &=& [i\,S\,,{\cal H}^{(0)}-i\,\partial_t],\nonumber \\
{\cal H}^{(j)} &=& \frac{1}{j}\,[i\,S\,,{\cal H}^{(j-1)}]\,, \ \ {\mbox{\rm for
    $j=2\ldots 6$}},
\label{b10}
\end{eqnarray}
and higher order terms with $j>6$ are
neglected. The calculation of nested commutators
is rather tedious but the result is simply
\begin{eqnarray} \label{HFWa}
H_{\rm FW} &=& e\,A^0 + \frac{(\bsigma\cdot\bpi)^2}{2\,m} - 
\frac{(\bsigma\cdot\bpi)^4}{8\,m^3}
+\frac{(\bsigma\cdot\bpi)^6}{16\,m^5}
\nonumber \\ && 
-\frac{i\,e}{8\,m^2}\,[\bsigma\cdot\bpi,\bsigma\cdot\bfE]
-\frac{e}{16\,m^3}\,\bigl\{\bpi\,,\,\partial_t{\bfE}\bigr\}
\nonumber \\ && 
-\frac{i\,e}{128\,m^4}\,[\bsigma\cdot\bpi,[\bsigma\cdot\bpi,
[\bsigma\cdot\bpi,\bsigma\cdot{\bfE}]]]
\nonumber \\ && 
+\frac{i\,e}{16\,m^4}\,\Bigl\{(\bsigma\cdot\bpi)^2\,,\,
[\bsigma\cdot\bpi,\bsigma\cdot\bfE]\Bigr\}\,.
\end{eqnarray}
There is some arbitrariness in the operator $S$, which means
that $H_{\rm FW}$ is not unique. The standard approach \cite{itzykson:80} relies
on the subsequent use of several FW transformations and yields a result that agrees with 
the $d=3$ limit of Eq.~(\ref{HFWa}) up to a transformation with an additional 
even operator.

Our aim is to obtain a Hamiltonian suitable for calculations
of the $m\,\alpha^7$ contributions to energy levels of an arbitrary light atom.
In this case one can neglect the vector potential $\bfA$ in all terms
having $m^4$ and $m^5$ in the denominator. Less obviously,
one can also neglect terms with  $\bsigma\cdot\bfA\,\bsigma\cdot\dot{\bfE}$ 
and $\bfB^2$. This is because they are of second order in
electromagnetic fields, which additionally contain derivatives, 
and thus contribute only to higher orders. 
After these simplifications, $H_{\rm FW}$ takes the form
\begin{eqnarray}
 H_{\rm FW} &=& e\,A^0 + \frac{\pi^2}{2\,m}-\frac{e}{4\,m}\,\sigma^{ij}\,B^{ij} - 
\frac{\pi^4}{8\,m^3}
\nonumber \\ &&
-\frac{e}{8\,m^2}\Bigl(\bnabla\cdot\bfE +
\sigma^{ij}\,\bigl\{E^i\,,\,\pi^j\bigr\}\Bigr)
\nonumber \\ &&
+\frac{e}{16\,m^3}\bigl\{\sigma^{ij} B^{ij}\,,\,p^2\bigr\}  
-\frac{e}{16\,m^3}\,\bigl\{\bfp\,,\, \partial_t{\bfE}\bigr\}
\nonumber \\ &&
+\frac{3\,e}{32\,m^4}\,\bigl\{\sigma^{ij}\,E^i\,p^j\,,\,p^2\bigr\}
+\frac{e}{128\,m^4}\,[p^2,[p^2,A^0]]
\nonumber \\ &&
-\frac{3\,e}{64\,m^4}\,\bigl\{p^2\,,\,\nabla^2 A^0 \bigr\}
+\frac{p^6}{16\,m^5}, \label{b09}
\end{eqnarray}
where
\begin{eqnarray}
\sigma^{ij} &=& \frac{1}{2\,i}\,[\sigma^i\,,\,\sigma^j],\\
B^{ij} &=& \partial^i\,A^j - \partial^j\,A^i,\\
E^i &=& -\nabla^i A^0 -\partial_t A^i.
\end{eqnarray}

\end{document}